\documentclass[]{pasj01}
\usepackage{natbib}
\usepackage{url}
\usepackage{txfonts}
\usepackage{xcolor}
\usepackage{listings} 
\usepackage{times}
\newcommand{\planck}{\textit{Planck}}  
\newcommand{\disperse}{{\tt  DisPerSE}}  
\newcommand{\herschel}{{\it Herschel}}
\def\nhunit{\ifmmode {\rm \,cm^{-2}} \else $\rm \,cm^{-2}$ \fi} % NH units
\def\nhh{\ifmmode N_{\rm H_{2}}\else $N_{\rm H_{2}}$\fi} 
\def\nhhc{\ifmmode N_{\rm H_{2}}^0\else $N_{\rm H_{2}}^0$\fi} 
\def\nhhbg{\ifmmode N_{\rm H_{2}}^{\rm bg}\else $N_{\rm H_{2}}^{\rm bg}$\fi} 
\def\ml{\ifmmode M_{\rm line}\else $M_{\rm line}$\fi}  
\def\sunpc{\ifmmode \rm M_\odot/\rm pc\else $\rm M_\odot/\rm pc$\fi}  
\def\rout{\ifmmode R_{\rm out}\else $R_{\rm out}$\fi}  
\def\av{\ifmmode A_{\rm V}\else $A_{\rm V}$\fi}   
\def\fwhmdec{\ifmmode FHWM_{\rm dec}\else $FWHM_{\rm dec}$\fi}   
\def\rflat{\ifmmode R_{\rm flat}\else $R_{\rm flat}$\fi}   
\def\kms{\ifmmode \,km\,s$^{-1}$\else \,km\,s$^{-1}$\fi} 
\def\arcm{\ifmmode {^{\scriptstyle\prime}}
          \else $^{\scriptstyle\prime}$\fi}
\newdimen\sa  \newdimen\sb
\def\parcs{\sa=.07em \sb=.03em
     \ifmmode \hbox{\rlap{.}}^{\scriptstyle\prime\kern -\sb\prime}\hbox{\kern -\sa}
     \else \rlap{.}$^{\scriptstyle\prime\kern -\sb\prime}$\kern -\sa\fi}
\def\parcm{\sa=.08em \sb=.03em
     \ifmmode \hbox{\rlap{.}\kern\sa}^{\scriptstyle\prime}\hbox{\kern-\sb}
     \else \rlap{.}\kern\sa$^{\scriptstyle\prime}$\kern-\sb\fi}
\def\rev{}
\begin{document} 
\Received{}%{yyyy/mm/dd}
\Accepted{}%{yyyy/mm/dd}
%\Published{yyyy/mm/dd}

\title{Molecular filament formation and  filament-cloud interaction: Hints from Nobeyama 45m telescope observations}

%%% begin:list of authors
% Do NOT capitalize all letters in "textsc".
\author{Doris \textsc{Arzoumanian}\altaffilmark{1}}%
%\thanks{Example: Present Address is xxxxxxxxxx}}
\altaffiltext{1}{Department of Physics, Graduate School of Science, Nagoya University, Furo-cho, Chikusa-ku, Nagoya 464-8602, Japan}
\email{doris.arzoumanian@nagoya-u.jp}
\author{Yoshito \textsc{Shimajiri}\altaffilmark{2}}
\altaffiltext{2}{Laboratoire AIM, CEA/DSM--CNRS--Universit\'e Paris Diderot, IRFU/Service d'Astrophysique, C.E.A. Saclay, Orme des Merisiers, 91191 Gif-sur-Yvette, France}
%\email{bbbbb@xxx.xxx.xx.xx}

\author{Shu-ichiro  \textsc{Inutsuka}\altaffilmark{1}}
\author{Tsuyoshi   \textsc{Inoue}\altaffilmark{1},}
\author{Kengo    \textsc{Tachihara}\altaffilmark{1}}
%\altaffiltext{3}{C-Address of Institute}
%\email{ccccc@xxx.xxx.xx.xx}
%%% end:list of authors

%% `\KeyWords{}' always has to be placed before `\maketitle'.
 %Do NOT move this preamble from here!
%\KeyWords{Stars: formation$_1$ --- ISM: clouds, filaments$_2$ --- Observations: dust and gas$_3$}

\maketitle

\begin{abstract}
We present Nobeyama 45m telescope C$^{18}$O, $^{13}$CO, and  $^{12}$CO$(1-0)$  mapping observations towards an interstellar filament in the Taurus molecular cloud. 
We investigate the gas velocity structure along the filament and in its surrounding parent cloud. The filament is detected in the optically thin C$^{18}$O emission as a single  velocity component, $\sim1\,$pc long, $\sim0.06\,$pc wide structure. The C$^{18}$O emission traces dust column densities larger than $\sim5\times10^{21}\nhunit$. The line-of-sight (LOS) velocity fluctuates along the filament crest with an average amplitude of  $\sim0.2\,$\kms. 
The $^{13}$CO and  $^{12}$CO integrated intensity maps show {\rev spatially extended %lower column density
 emission around the %denser 
 elongated filament.} 
We identify  three extended structures with LOS velocities redshifted and blueshifted with respect to the average velocity of the filament identified in C$^{18}$O.
Based on  combined analyses of velocity integrated channel maps and intensity variations of the optically thick  $^{12}$CO spectra on and off the filament, we propose a 3-dimensional structure of the cloud surrounding the filament.  
We further suggest a multi-interaction scenario where sheet-like extended structures interact, in space and time, with the filament and are responsible for its compression and/or disruption, playing an important role in the star formation history of the filament.
We also identify, towards the same field, a very faint filament showing a velocity field compatible with the filament formation process  proposed by \citet{Inoue2017}, where a filament is formed 
due to convergence of a flow of matter generated by the bending of the ambient magnetic field structure induced by an interstellar shock compression. 
\end{abstract}

\section{Introduction}

\begin{figure*}[!h]
 \begin{center}
 \hspace{-0.2cm}
  \includegraphics[width=6.cm]{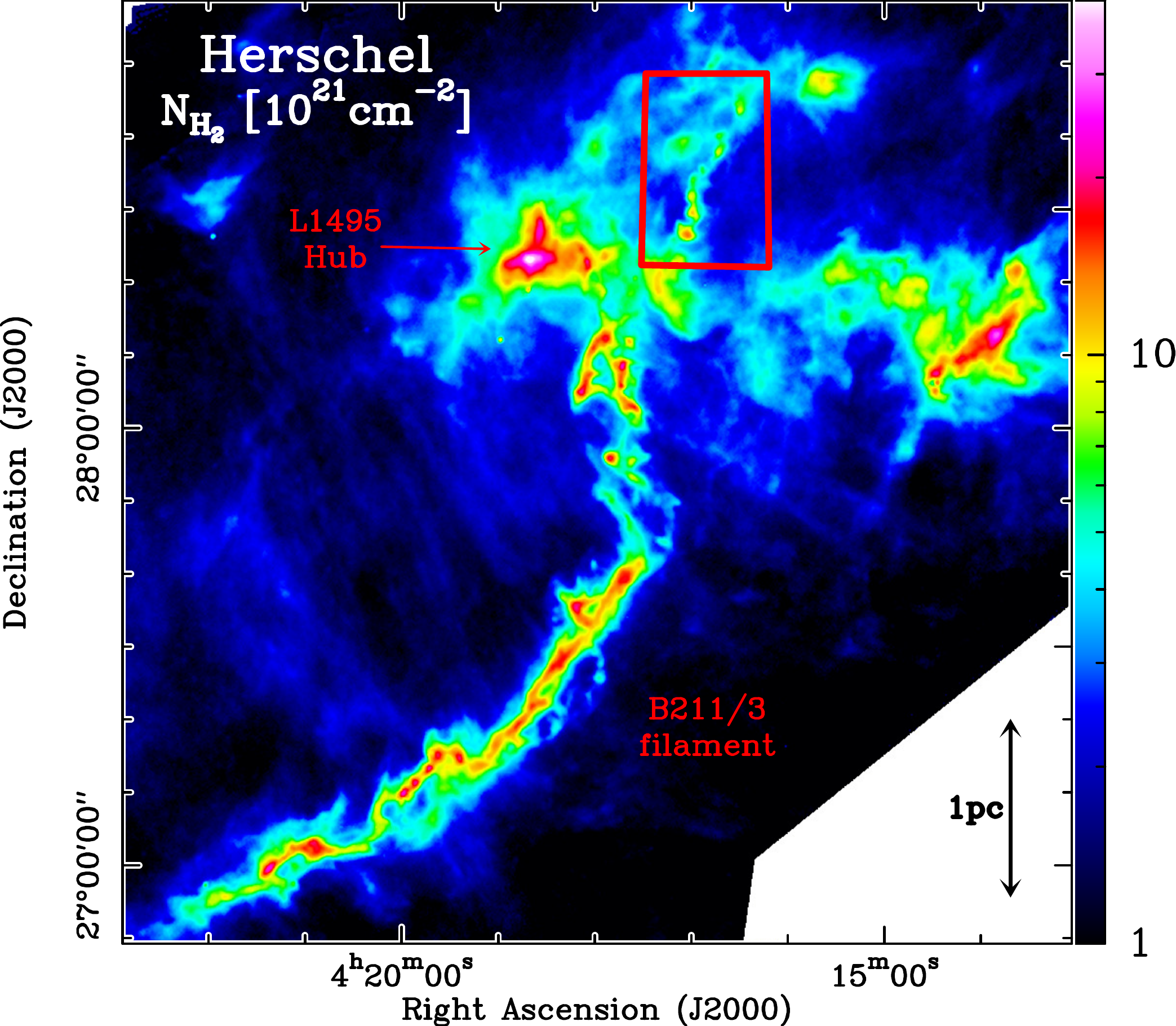} %4.9
 \hspace{0.3cm}
  \includegraphics[width=4.5cm]{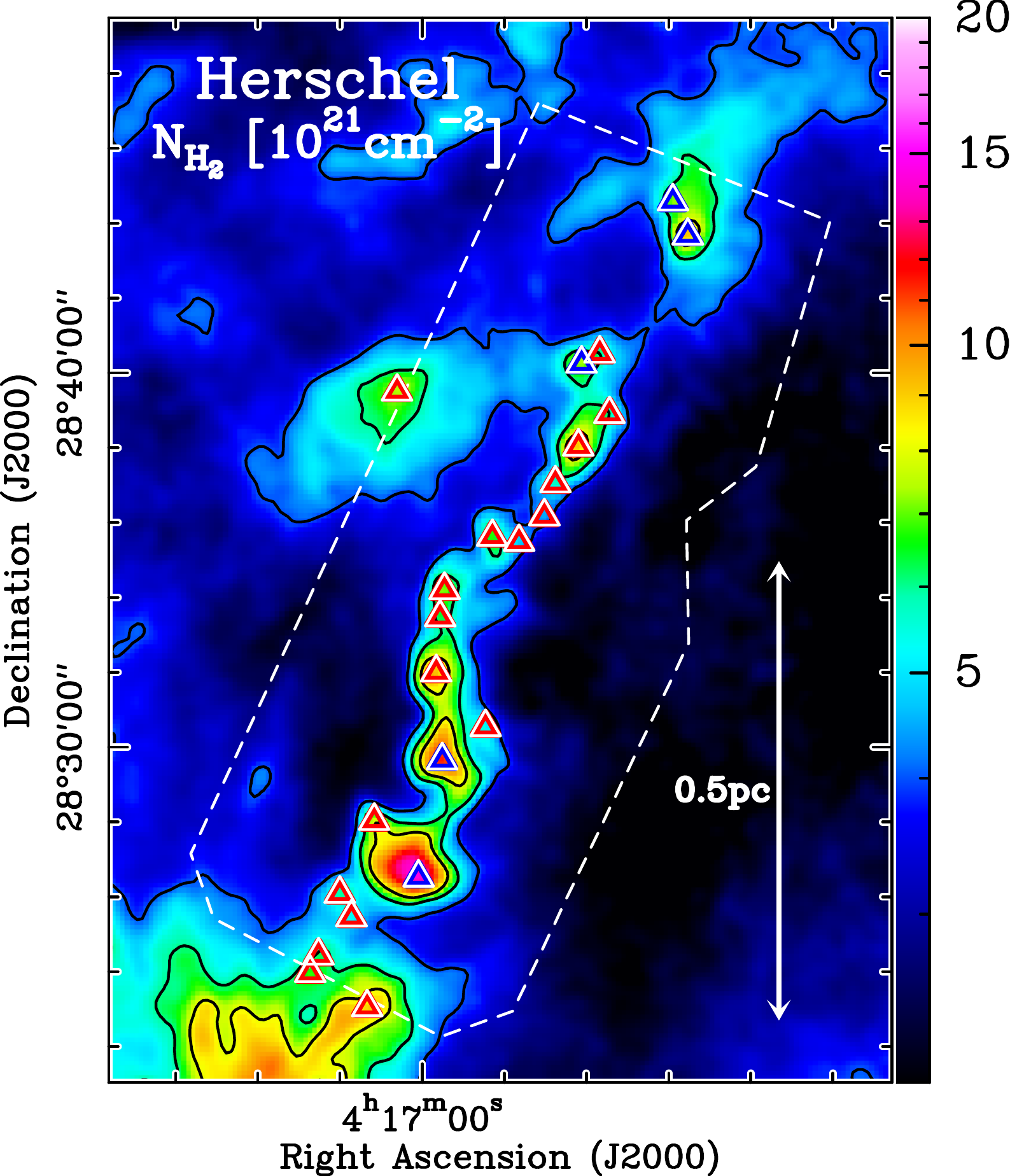}  % 3.7
   \hspace{0.3cm}
  \includegraphics[width=4.5cm]{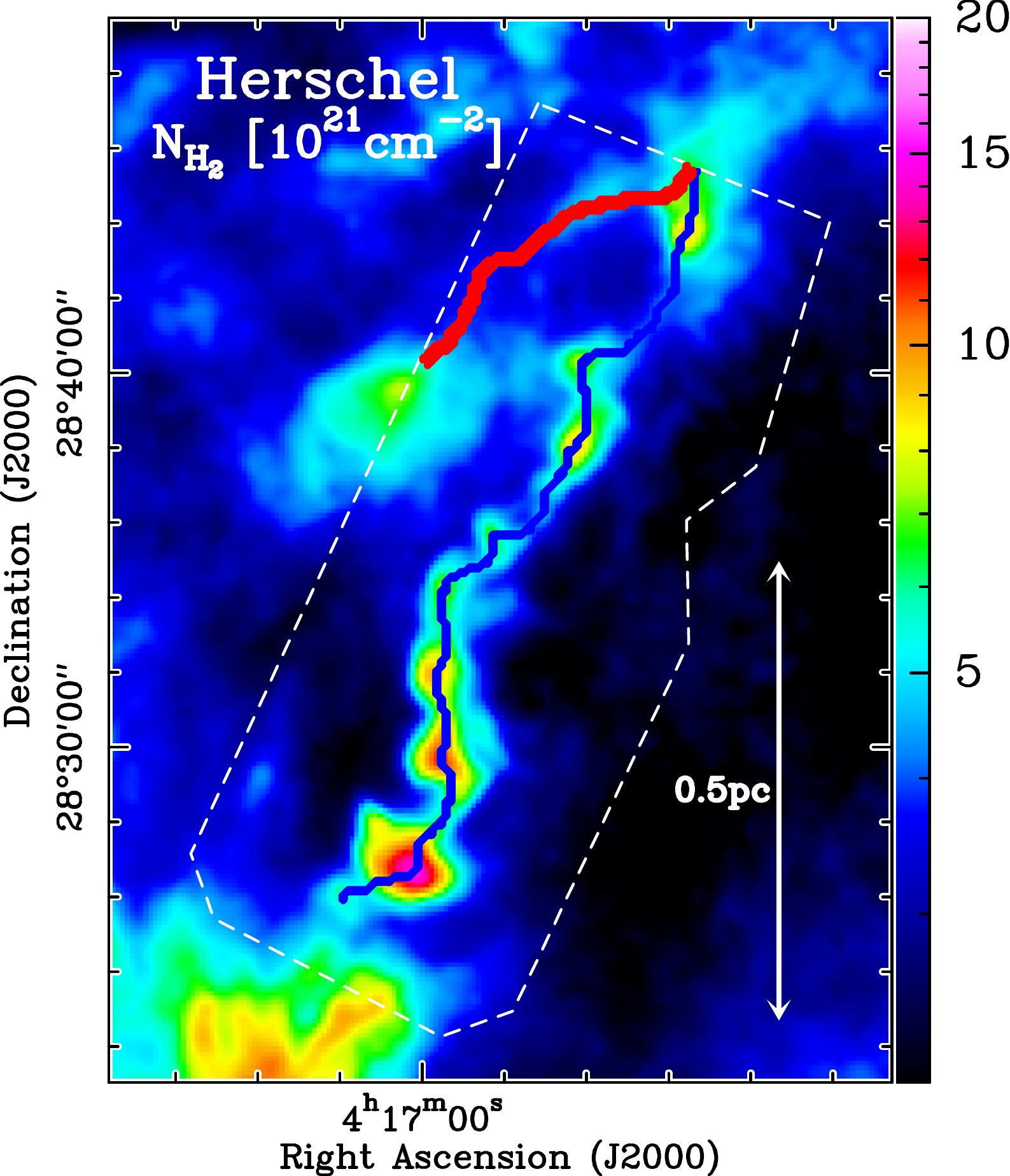} 
 \end{center}
\caption{{\it Left}:   Column density map of part of the Taurus molecular cloud 
as derived from HGBS data 
  \citep[http://gouldbelt-herschel.cea.fr/archives -- cf.][]{Palmeirim2013}. The B211/3 well studied filament and the L1495 star forming hub are indicated. The red rectangle shows the region analyzed in this paper.   
	        {\it Middle}:   Blow up column density map, at the resolution of 28", of the red rectangle plotted on the  {\it left} panel.  The black contours are at column densities of 4, 6, and 8\,$\times10^{21}\,\nhunit$. 
	        The blue- and red-white triangles are the prestellar cores and the unbound fragments identified by  \citet{Marsh2016}, respectively. 
	        The white dashed contour shows the footprint of the CO maps observed with the Nobeyama 45m telescope.
 {\it Right}: The map is the same as the middle panel. The blue and red skeletons trace the crests of the two filaments discussed in this paper, {\bf MF} and {\bf YF}, respectively. 
}
	        \label{Herschel_map}
\end{figure*}

Molecular clouds are observed to be filamentary 
\citep[e.g.,][]{Andre2010,Molinari2010,Umemoto2017}.   
Filamentary molecular clouds are proposed to be formed out of dense and cold atomic clouds as a result of multiple compressions from propagating shock waves through the interstellar medium \citep[e.g.,][]{Hennebelle2008,Inoue2009,Inutsuka2015}.
The typical timescale of such shock compressions is estimated to be on average $\sim1$\,Myr \citep{McKee1977}. After each passage of a wave, the properties of the shocked molecular interstellar medium (ISM) are modified. Thus   the present morphologies  of the density, velocity, and magnetic field structures, are usually not those corresponding to the initial conditions at the molecular cloud formation epoch, but are probably the result of their sequential alteration due to interactions with multiple propagating ISM waves. 

Hence, in order to describe the formation and evolution of  structures  in molecular clouds, one should take into account the reorganization of interstellar  matter, as a function of time, due to the propagation of waves through the filamentary clouds.

In the context of star formation, these propagating shock waves, may as a consequence be responsible for the 
assembly of dense molecular matter  in  the form of ``thermally supercritical''  filaments where the bulk of star formation  is observed to take place \citep[e.g.,][]{Andre2010,Konyves2015,Marsh2016}. These thermally supercritical filaments are characterized with  a mass per unit length, $M_{\rm line}$, of the order or larger than the critical line mass of nearly isothermal, long cylinders,  $M_{\rm line,crit}=2 c_{\rm s}^{2}/G \sim16\,\sunpc$ \citep[][where $c_{\rm s} \sim0.2$\,km\,s$^{-1}$  for gas temperature of $ \sim 10$\,K]{Stodolkiewicz1963,Ostriker1964}, and are unstable for radial collapse and fragmentation  \citep[cf.][]{Inutsuka1997}. This critical mass per unit length corresponds to a  column density of  $\nhh\sim8\times10^{21}$~cm$^{-2}$ or a surface density of   $\Sigma_{\rm fil}^0\sim116\,M_\odot /{\rm pc}^2$ for 0.1\,pc wide filaments \citep[][]{Arzoumanian2011}. 
These latter values are comparable to the  proposed 
column density threshold for star formation  \citep{Lada2010,Shimajiri2017}. Hence, the  column density threshold for star formation  can  now be understood as the  threshold in filament $M_{\rm line}$ equivalent to a critical value of hydrostatical equilibrium above which filaments are unstable to radial collapse and fragmentation into star forming cores \citep[cf.][]{Andre2014}.

The analyses of \herschel\ observations towards the Aquila star forming region suggest that a relatively small fraction of the matter, about $15\,\%$ on average, of the mass of supercritical filaments is in the form of prestellar cores \citep{Konyves2015}. Understanding this observed fraction, in the light of star formation along supercritical filaments, may give us a hint on the origin of the observed star formation efficiency in molecular clouds and in the Galaxy. 

 Molecular filaments are observed to span a wide range in central column density, mass per unit length, length, while they all  share the same central width of about 0.1pc \citep[as derived in nearby regions from \herschel\ dust continuum observations,][sub., and others]{Arzoumanian2011,JuvelaMalinen2012,AlvesdeOliveira2014,Koch2015,Arzoumanian2018}.   The origin of this filament property is not yet understood. 
\citet[][]{Arzoumanian2011} suggested that the characteristic filament width may be linked to the sonic scale of turbulence in the cold ($\sim10$\,K) ISM, observed to be around 0.1\,pc \citep[][]{Larson1981,Goodman1998}. 
The latter appears to be roughly the scale at which supersonic magnetohydrodynamic (MHD) turbulence dissipates \citep[][]{Federrath2010,Vazquez-Semadeni2003}, compatible 
 with the observed subsonic to transonic velocity dispersions of subcritical/critical filaments \citep{Arzoumanian2013,Hacar2013,Hacar2016}. 
These results suggest that the dissipation of large-scale shock waves  in the ISM 
may be important for the formation of the observed filamentary web  \citep[][]{Inutsuka2015,Inoue2017}.

\begin{figure*}[!h]
 \begin{center}
 \hspace{-0.2cm}
  \includegraphics[width=5.5cm]{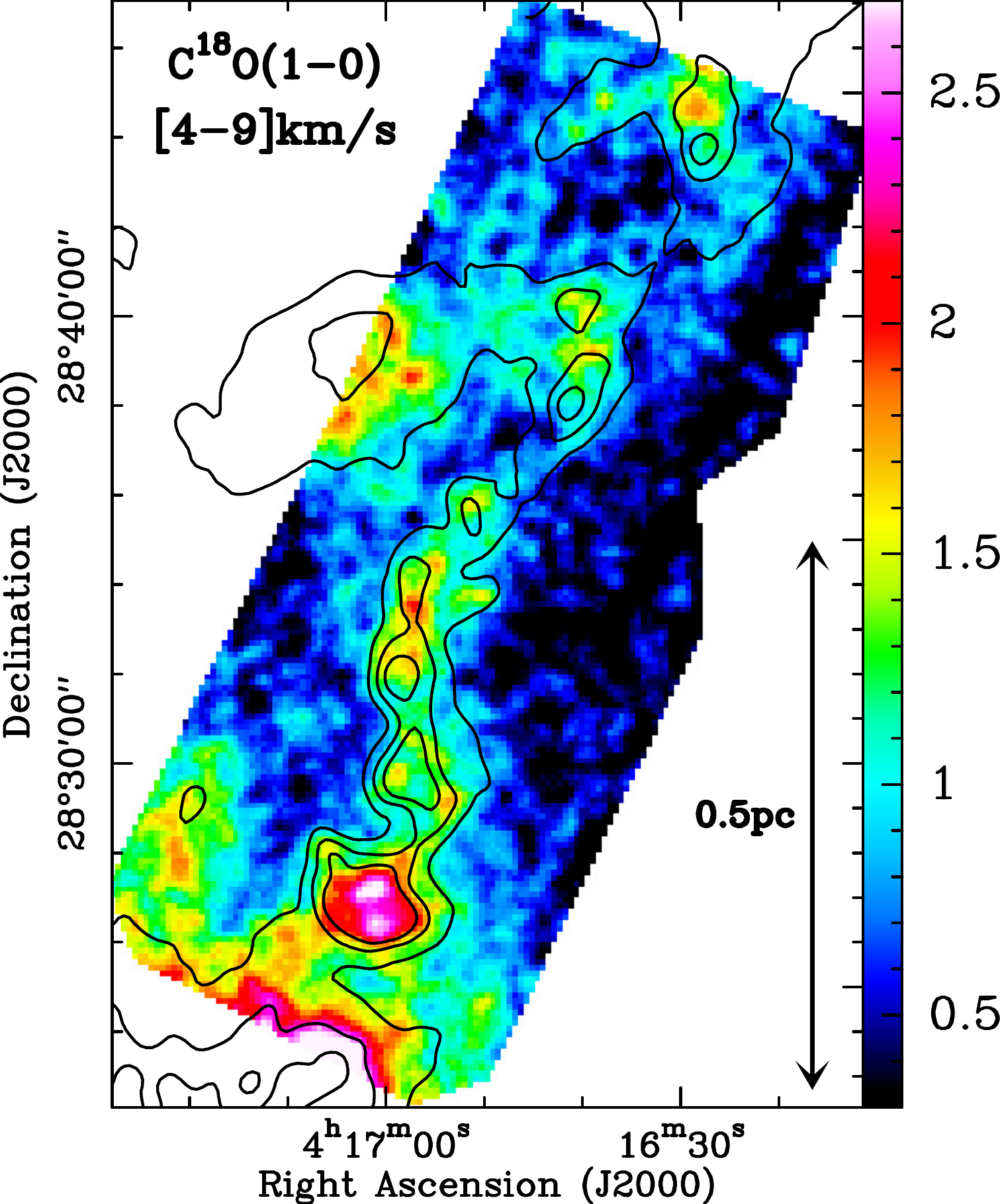} 
   \hspace{0.2cm}
  \includegraphics[width=5.5cm]{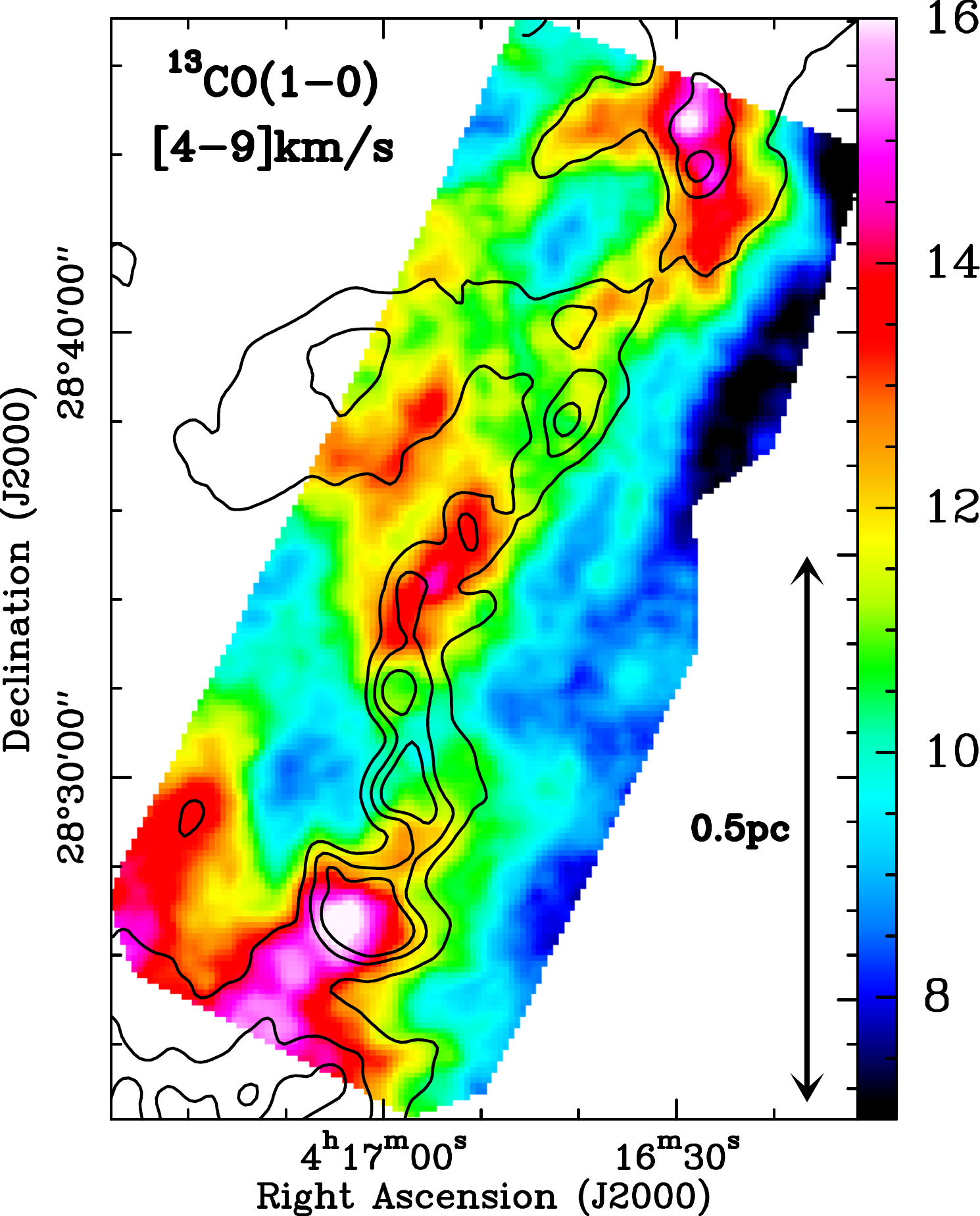} 
   \hspace{0.2cm}
  \includegraphics[width=5.5cm]{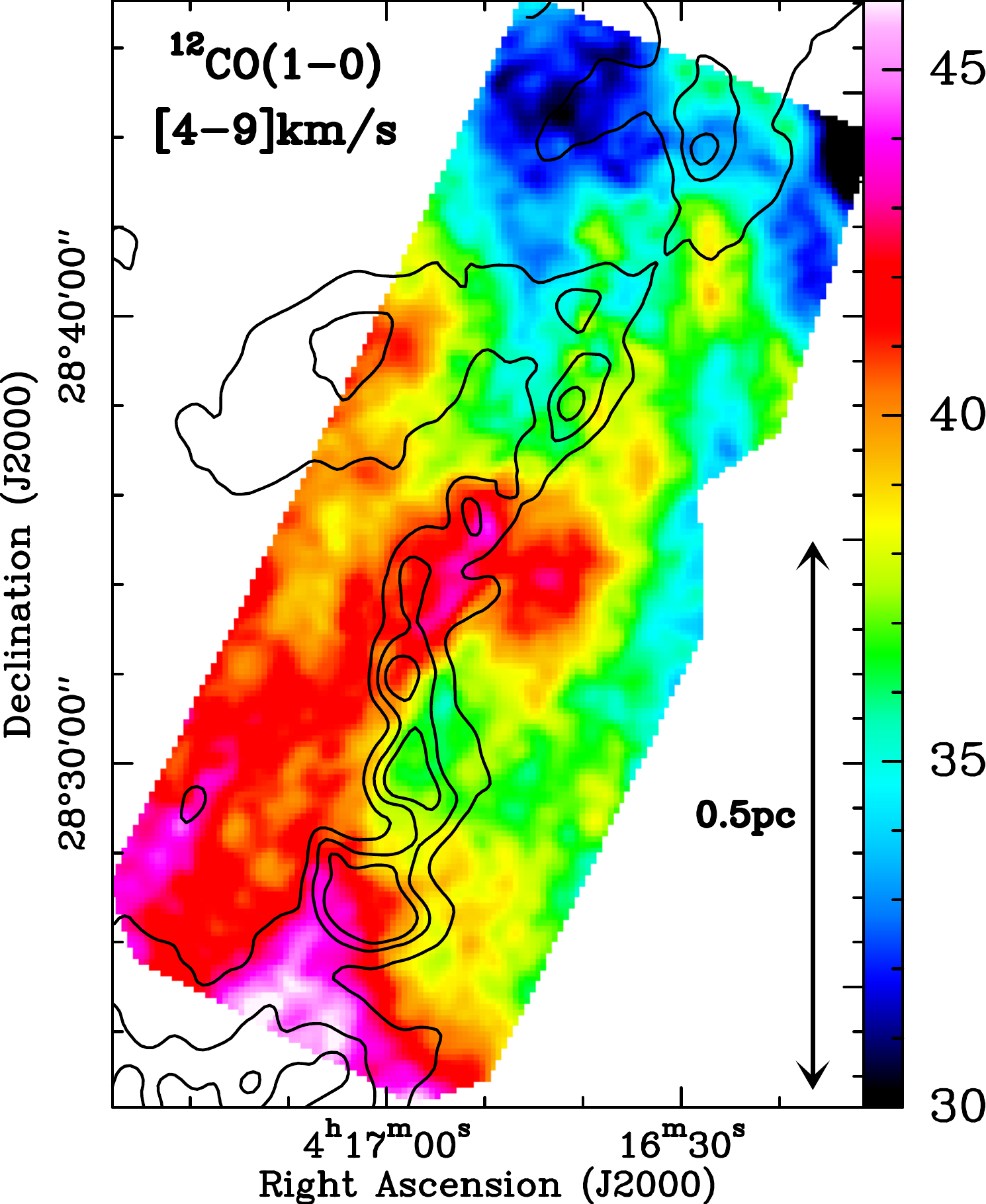} 
 \end{center}
\caption{From left to right, C$^{18}$O$(1-0)$, $^{13}$CO$(1-0)$, and  $^{12}$CO$(1-0)$  Integrated intensity maps (in units of K\,kms$^{-1}$) over the LSR velocity range 4 to 9\,kms$^{-1}$. The spatial and spectral resolutions of the maps are $28\arcsec$ and 0.07\,\kms, respectively.
The black contours correspond to column densities of 4, 6, and 8\,$\times10^{21}\,\nhunit$ derived from \herschel\ data, and are the same as those in Fig.\,\ref{Herschel_map}. The C$^{18}$O$(1-0)$ integrated emission traces the elongated structure of the filament.
}
	        \label{CO_integratedMaps}
\end{figure*}

In this paper we present Nobeyama 45m telescope C$^{18}$O$(1-0)$, $^{13}$CO$(1-0)$, and  $^{12}$CO$(1-0)$ molecular line observations towards a prominent $\sim1$\,pc long filament identified with \herschel\ in the Taurus molecular cloud (Fig.\,\ref{Herschel_map}). In Sect.\,\ref{ObsData}, we present our mapping observations with the Nobeyama 45m telescope. Section\,\ref{AnaResults} details the results derived from the analysis of the  velocity cubes, the spectra, the channel maps, the position-velocity maps,  along, across and around the filament. In Sect.\, \ref{Discussion}, we discuss the implication of our results in the understanding of filament formation, evolution, and interaction with the surrounding cloud. We summarize the results presented in this paper and conclude in Sect.\,\ref{Summary}. Three appendices %(\ref{AppSimu}, \ref{AppSpectra}, and \ref{HerschelObs}) 
complement the analyses presented in the main text.

\section{Molecular line mapping observations with the Nobeyama 45m telescope}\label{ObsData}

We used the Nobeyama 45m telescope to map  a molecular filament previously identified by \herschel, and its surrounding parent cloud (see Fig.\,\ref{Herschel_map}). The target region is  located on the west of the star forming L1495 hub in the Taurus molecular clouds at a distance of 140\,pc \citep[][]{Elias1978,Myers2009,Ward-Thompson2016}. 
The observations were carried out 
in Februrary 2017. 
We mapped a 0.14 deg$^2$ region around the filament  in $^{12}$CO$(1-0)$, $^{13}$CO$(1-0)$, and C$^{18}$O$(1-0)$
  with the FOREST receiver \citep{Minamidani2016}.  All molecular line data were obtained simultaneously. At 115 GHz, the telescope has a beam size of $\sim15\arcsec$ (HPBW). 
 As backend, we used the SAM45 spectrometer which provides a bandwidth of 31 MHz and a frequency resolution of 7.63 kHz. The latter corresponds to a velocity resolution of $\sim$0.02 km s$^{-1}$ at 115 GHz. The standard chopper wheel method was used to convert the observed signal to the antenna temperature $T_{\rm A}^*$ in units of K, corrected for the atmospheric attenuation. %
To estimate the main beam brightness temperature, $T_{\rm MB}$,   
we mapped a small area of the OMC-2/FIR 4 region with strong dust continuum emission in the OMC-2/3 region \citep{Shimajiri2008,Shimajiri2015a,Shimajiri2015b} once or twice per observing day. We scaled the FOREST intensity ($T_{\rm A}^*$) to the intensity of the BEARS data ($T_{\rm MB}$) obtained in  \citep{Shimajiri2011,Shimajiri2014} by comparing the FOREST intensity with the BEARS intensity in the OMC-2/FIR 4 region.
 During the observations, the system noise temperatures ranged from 150\,K to 580\,K. The telescope pointing was checked every hour by observing the SiO maser source  NML-tau, and was better than 3$\arcsec$ throughout the entire observing run. 
 We used the on-the-fly (OTF) mapping technique. The central position of our final map is: (RA$_{\rm J2000}$, DEC$_{\rm J2000}$) = (04:16:49.711, 28:35:42.35).   We have chosen line-of-sights of (RA$_{\rm J2000}$, DEC$_{\rm J2000}$) = (04:11:48.691, 26:46:25.59) as the off position for baseline removal. 
We obtained OTF maps with two different scanning directions along the RA and Dec axes and combined them into a single map (using the \citet{Emerson1988} PLAIT algorithm) to reduce the scanning effects. We then smoothed spatially and spectrally all the three cubes  
to the same effective HPBW size of 28$\arcsec$ and velocity resolution of 0.07\,km\,s$^{-1}$. 
The 1$\sigma$ noise level of the final data at 28$\arcsec$  and 0.07\,km\,s$^{-1}$ are 0.48\,K, 0.22\,K, and 0.20\,K in $T_{\rm MB}$
for $^{12}$CO$(1-0)$, $^{13}$CO$(1-0)$, and C$^{18}$O$(1-0)$, respectively.

 \begin{figure}
 \begin{center}
  \includegraphics[width=8cm]{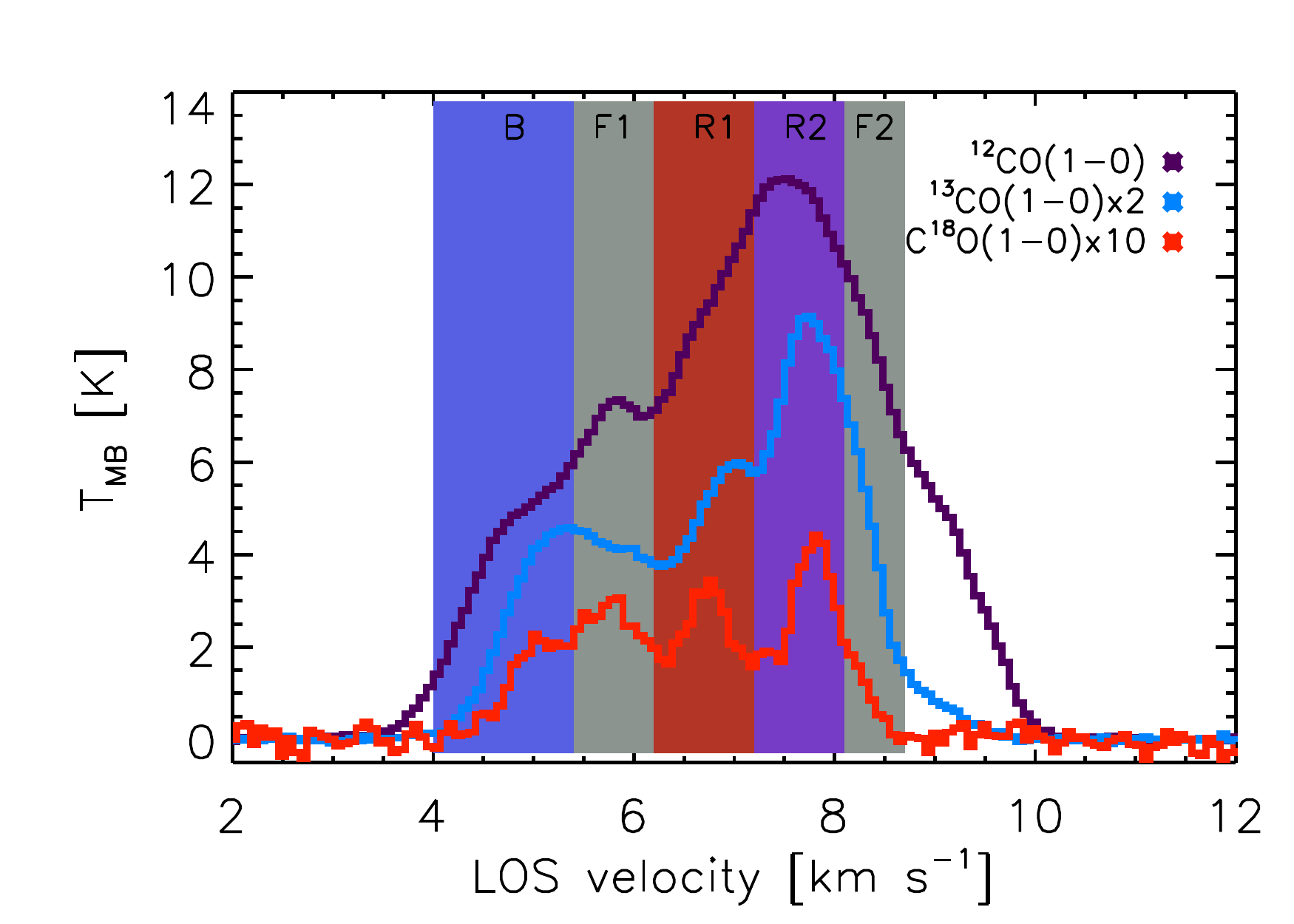} 
 \end{center}
\caption{Positionally averaged spectra (across the whole observed region)  of $^{12}$CO$(1-0)$, $^{13}$CO$(1-0)$, and C$^{18}$O$(1-0)$ emission. The $^{13}$CO and C$^{18}$O  spectra are multiplied by a factor 2 and 10, respectively.
The five colored vertical rectangles show the five velocity channels, which we use to divide the total integrated emission based on position-velocity diagrams as explained in Sect.\,\ref{PVmaps}. The Five channels are as follow:  
 {\bf B}[4-5.4]\kms, {\bf F1}[5.4-6.2]\kms, {\bf R1}[6.2-7.2]\kms, {\bf R2}[7.2-8.1]\kms, {\bf F2}[8.1-8.7]\kms.  
}
	        \label{CO_All_spectra}
\end{figure}

  \begin{figure}[!h]
 \begin{center}
 \hspace{-0.1cm}
  \includegraphics[width=3.9cm]{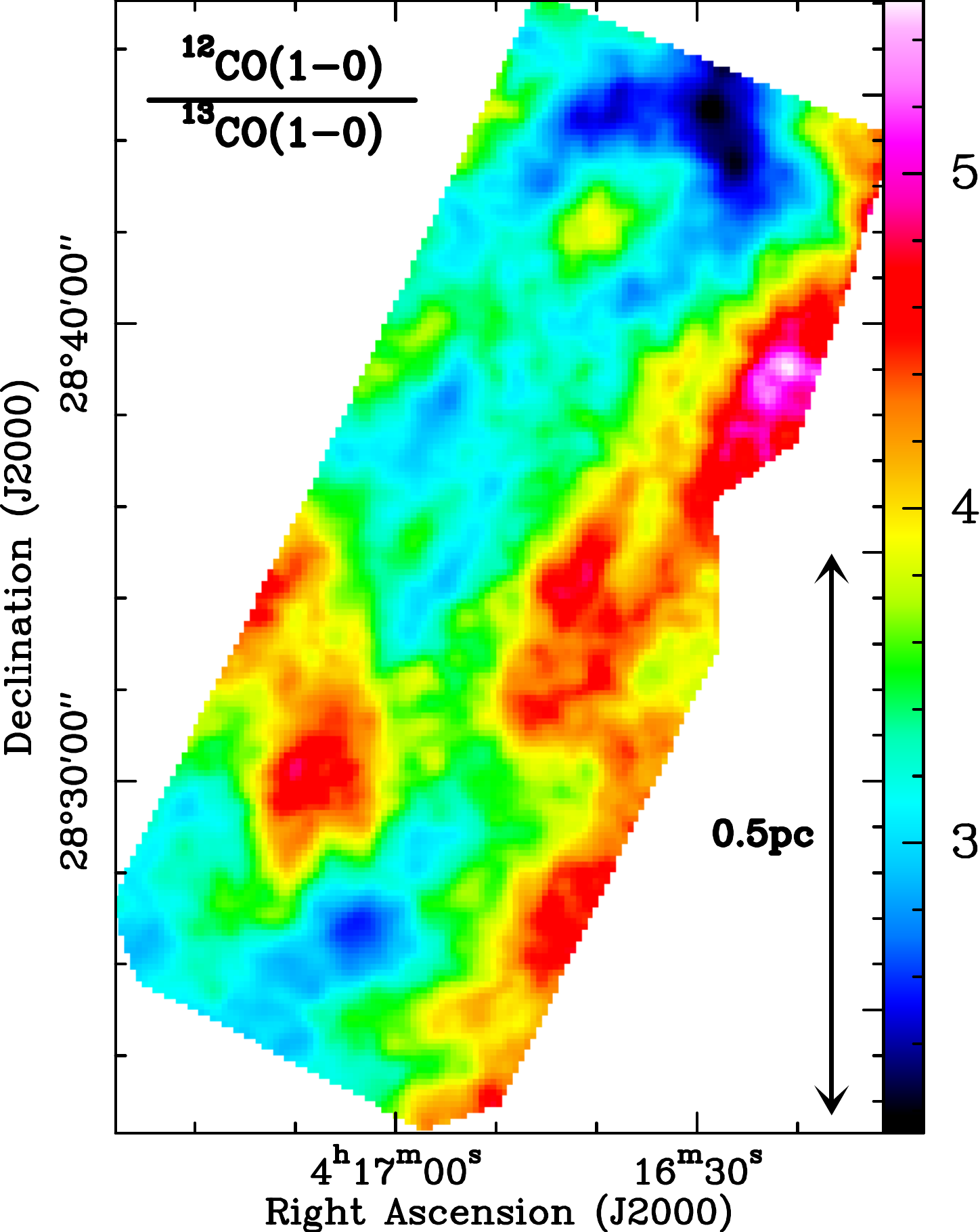} 
   \hspace{0.1cm}
  \includegraphics[width=4.cm]{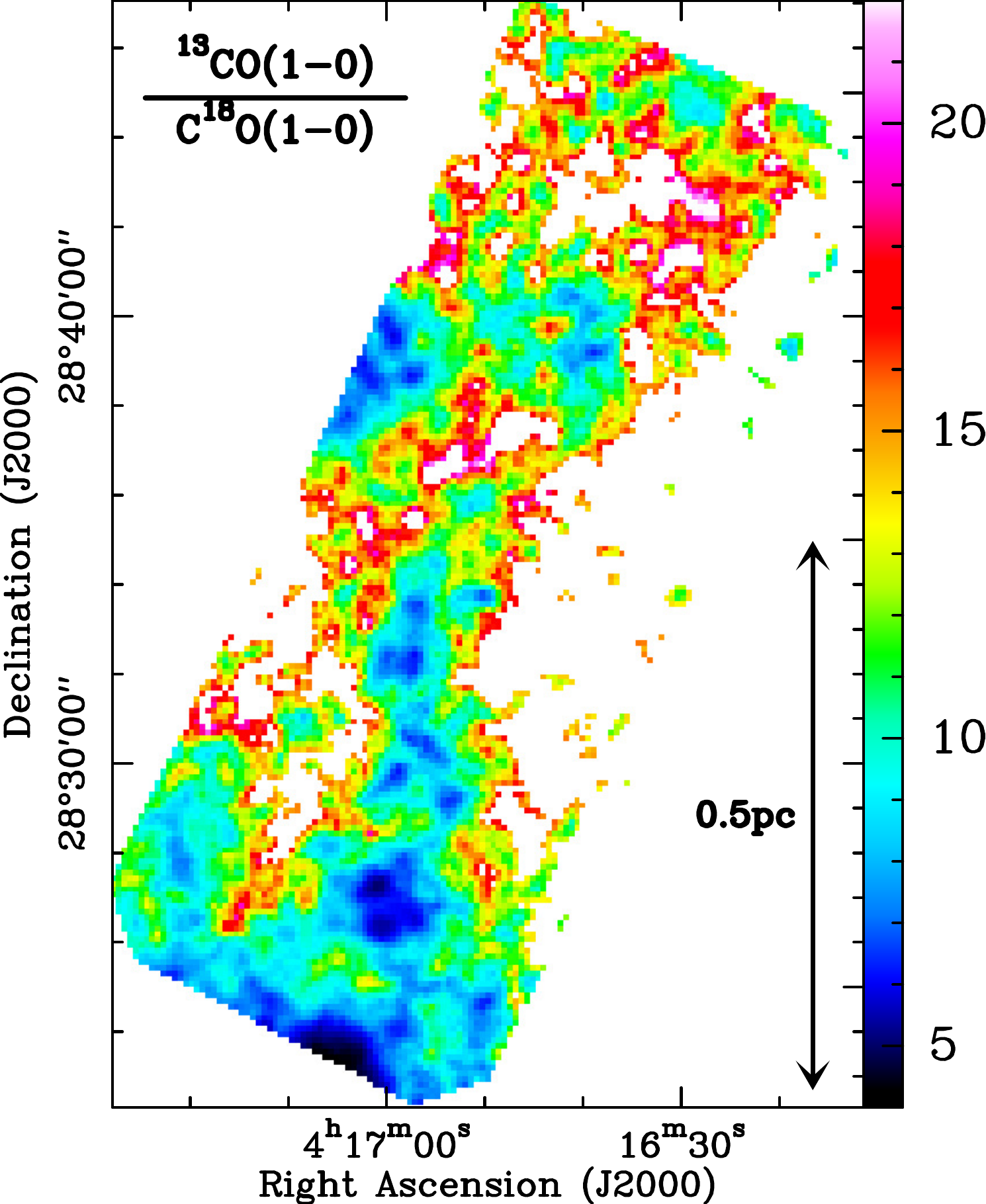} 
 \end{center}
\caption{Integrated intensity ratio map $\int_{4}^{9}T_{\rm MB}^{12}(\upsilonup ){\rm d}{\it \upsilonup }/ \int_{4}^{9} T_{\rm MB}^{13}({\it \upsilonup }){\rm d}{\it \upsilonup }$ (left)  and $\int_{4}^{9}T_{\rm MB}^{13}(\upsilonup ){\rm d}{\it \upsilonup }/ \int_{4}^{9} T_{\rm MB}^{18}({\it \upsilonup })\rm d{\it \upsilonup }$ (right). 
The $T_{\rm MB}^{12}$, $T_{\rm MB}^{13}$, and $T_{\rm MB}^{18}$ correspond to the mean beam brightness temperature of the $^{12}$CO$(1-0)$,  $^{13}$CO$(1-0)$, and C$^{18}$O$(1-0)$ emission, respectively,  $\upsilonup$ is the LOS velocity in the range [4-9]\,\kms. 
The integrated intensity maps are the same as in Fig.\,\ref{CO_integratedMaps}. For the right hand side map, only pixels 
with $T_{\rm MB}^{18}>5\sigma$  have been considered, where $\sigma=0.2$\,K (Sect.\,\ref{ObsData}). The $^{12}$CO$(1-0)$  emission is optically thick all over the observed field. The C$^{18}$O$(1-0)$ and $^{13}$CO$(1-0)$ emission are optically thin. 
}
	        \label{CO_integratedMaps_ratio}
\end{figure}

    \begin{figure}
 \begin{center}
 \hspace{-0.7cm}
  \includegraphics[width=8.8cm]{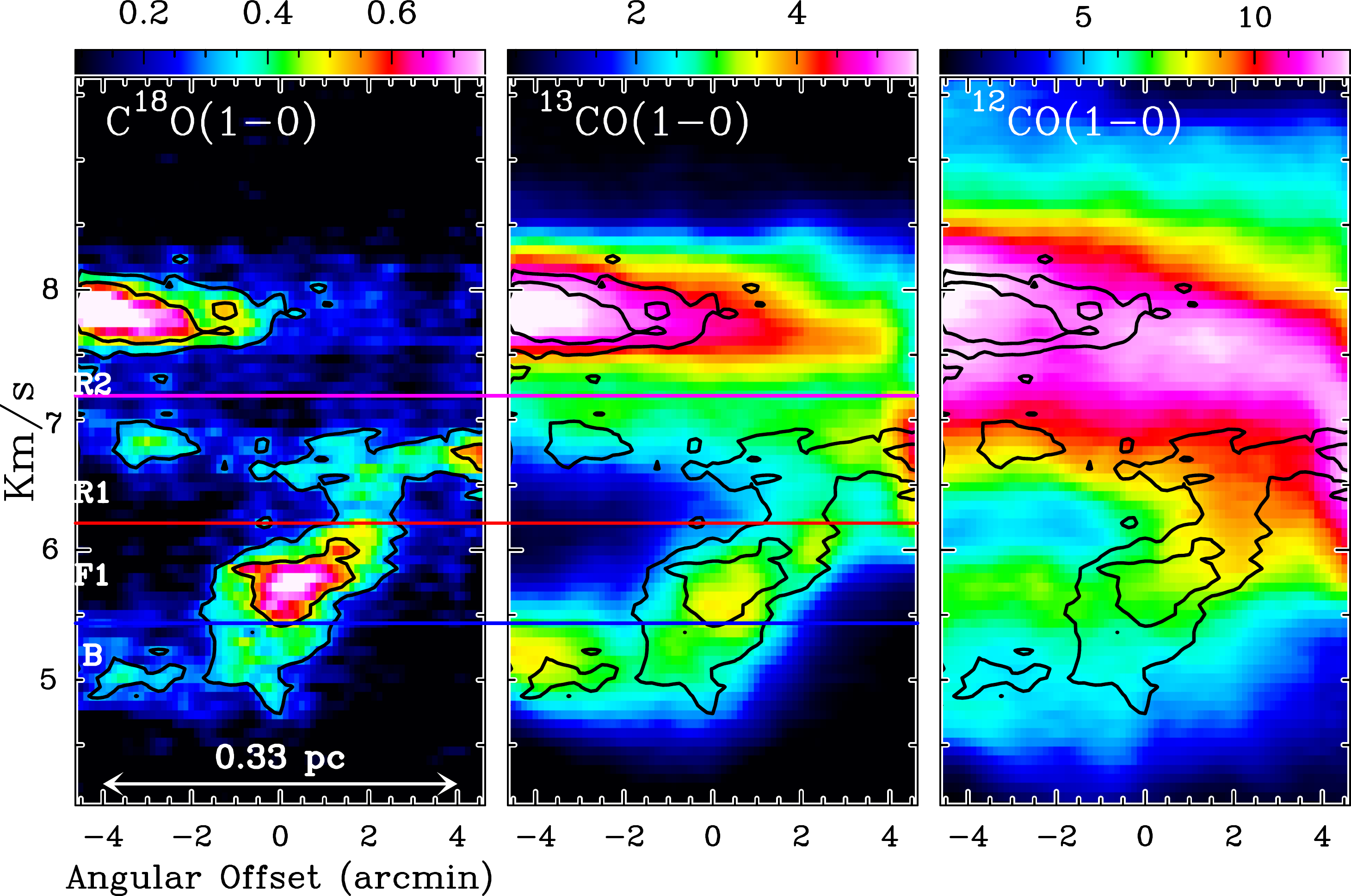} 
 \end{center}
\caption{     
Position velocity (PV) diagrams, in units of K($T_{\rm MB}$),  of the C$^{18}$O$(1-0)$, $^{13}$CO$(1-0)$, and $^{12}$CO$(1-0)$ emission, from left to right. These PV diagrams are perpendicular to the {\bf MF} filament axis and  averaged along its crest. The zero offset position (in the x-axis) indicates the position of the filament. The black contours correspond to the C$^{18}$O$(1-0)$ intensity at  0.3  and 0.5\,K, and are the same for the three panels.   The three horizontal lines, in blue, red, and purple show  velocities of 5.4, 6.2, and 7.2\,\kms, respectively.  These lines identify the four velocity ranges,  {\bf B},  {\bf F1},  {\bf R1},  and {\bf R2} described in Sect.\,\ref{PVmaps}.
}
	        \label{CO_PV}
\end{figure}

\section{Analyses and results}\label{AnaResults}

\subsection{Integrated intensity maps of $^{12}$CO$(1-0)$, $^{13}$CO$(1-0)$ and C$^{18}$O$(1-0)$}

We derived velocity  integrated intensity maps for the three lines of the observed region (Fig.\,\ref{CO_integratedMaps}). The emission is integrated for the local standard of rest (LSR) velocity range between 4 and 9\,km\,s$^{-1}$, which encompasses the bulk of the emission of the observed region, as can be seen on the   $^{12}$CO$(1-0)$, $^{13}$CO$(1-0)$ and C$^{18}$O$(1-0)$ spectra  averaged across the whole field (Fig.\,\ref{CO_All_spectra}). 

The C$^{18}$O$(1-0)$ integrated emission traces the elongated structure of the filament (Fig.\,\ref{CO_integratedMaps}) with column densities $\nhh\gtrsim5\,\times10^{21}\,\nhunit$ as derived from \herschel\ data (Fig.\,\ref{Herschel_map}, see also Appendix\,\ref{HerschelObs}). %On the other hand 
{\rev The $^{13}$CO and $^{12}$CO integrated intensity maps  show more extended emission around the filament  for column densities  $\nhh\gtrsim1\,\times10^{21}\,\nhunit$}.
%between $\sim1$\,and\,4$\times10^{21}\,\nhunit$. 
%From now on we call 
{\rev In the following we refer to} 
the filament identified with \herschel\ and traced with the C$^{18}$O$(1-0)$ emission as the main filament, {\bf MF}. 
The filament {\bf MF} can hardly be recognized on the velocity integrated map of the $^{13}$CO emission and cannot be seen in the $^{12}$CO emission map (Fig.\,\ref{CO_integratedMaps}).

We derive  $^{12}$CO$(1-0)$/$^{13}$CO$(1-0)$ and $^{13}$CO$(1-0)$/C$^{18}$O$(1-0)$ ratio maps, from the velocity integrated maps over the LSR velocity range 4 to 8.7\,km\,s$^{-1}$  (Fig.\,\ref{CO_integratedMaps}),  to estimate the mean optical depth of the three  lines \citep[cf., e.g., Sect.\,4 of][]{Arzoumanian2013}. The  mean optical depth values  of the $^{12}$CO$(1-0)$, $^{13}$CO$(1-0)$, and C$^{18}$O$(1-0)$ lines {\rev for the integrated emission} are estimated to be  between $ 12 < \tau_{^{12}\rm{CO}} < 39$, $ 0.06 < \tau_{^{13}\rm{CO}} < 0.9$, and $0.01 < \tau_{\rm{C}^{18}\rm{O}} <0.15$, respectively,  assuming mean values of the abundance ratios  [$^{12}$CO]/[$^{13}$CO]=62 \citep[][]{Langer1993} and [$^{13}$CO]/[C$^{18}$O]=5.5 \citep[][]{Wilson1994} in the local ISM (see also Table\,\ref{tab:coldens_tau}).
 The $^{12}$CO$(1-0)$  emission is found to be  highly optically thick all over the observed field, while the C$^{18}$O$(1-0)$ and $^{13}$CO$(1-0)$ emission are mostly optically thin. 
 {\rev 
Table\,\ref{tab:coldens_tau} below gives the optical depth values for 
%This result is also valid if we consider 
the five channel maps derived by integrating the emission over smaller velocity ranges corresponding to the five identified emission structures with different velocity components. While the C$^{18}$O emission is optically thin for all channels, the  $^{13}$CO emission is estimated to be locally slightly optically thick  (cf. Sect.\,\ref{PVmaps} below and Table\,\ref{tab:coldens_tau}).}

  \begin{figure*}
 \begin{center}
  \resizebox{15cm}{!}{
   \includegraphics[]{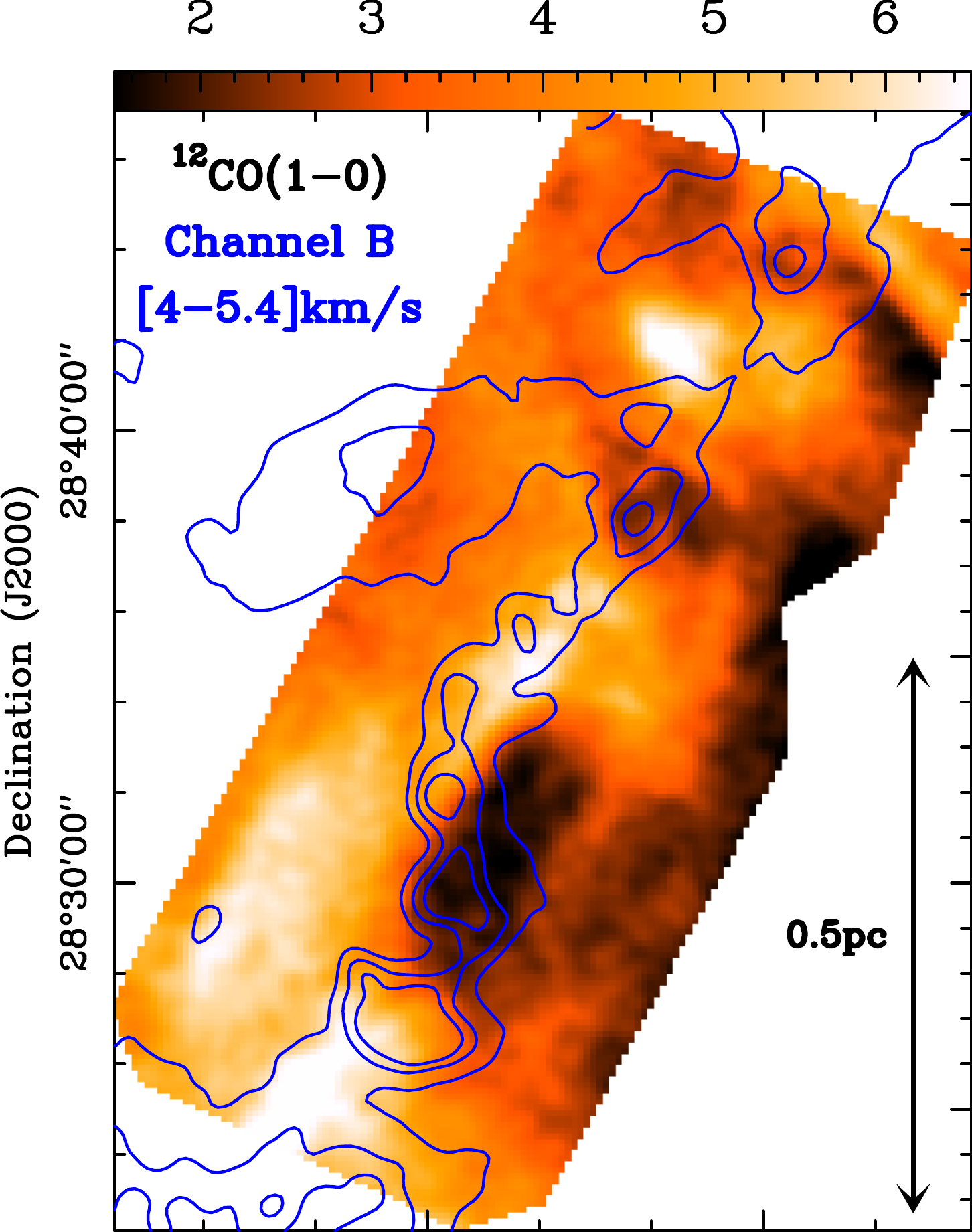} 
  \includegraphics[]{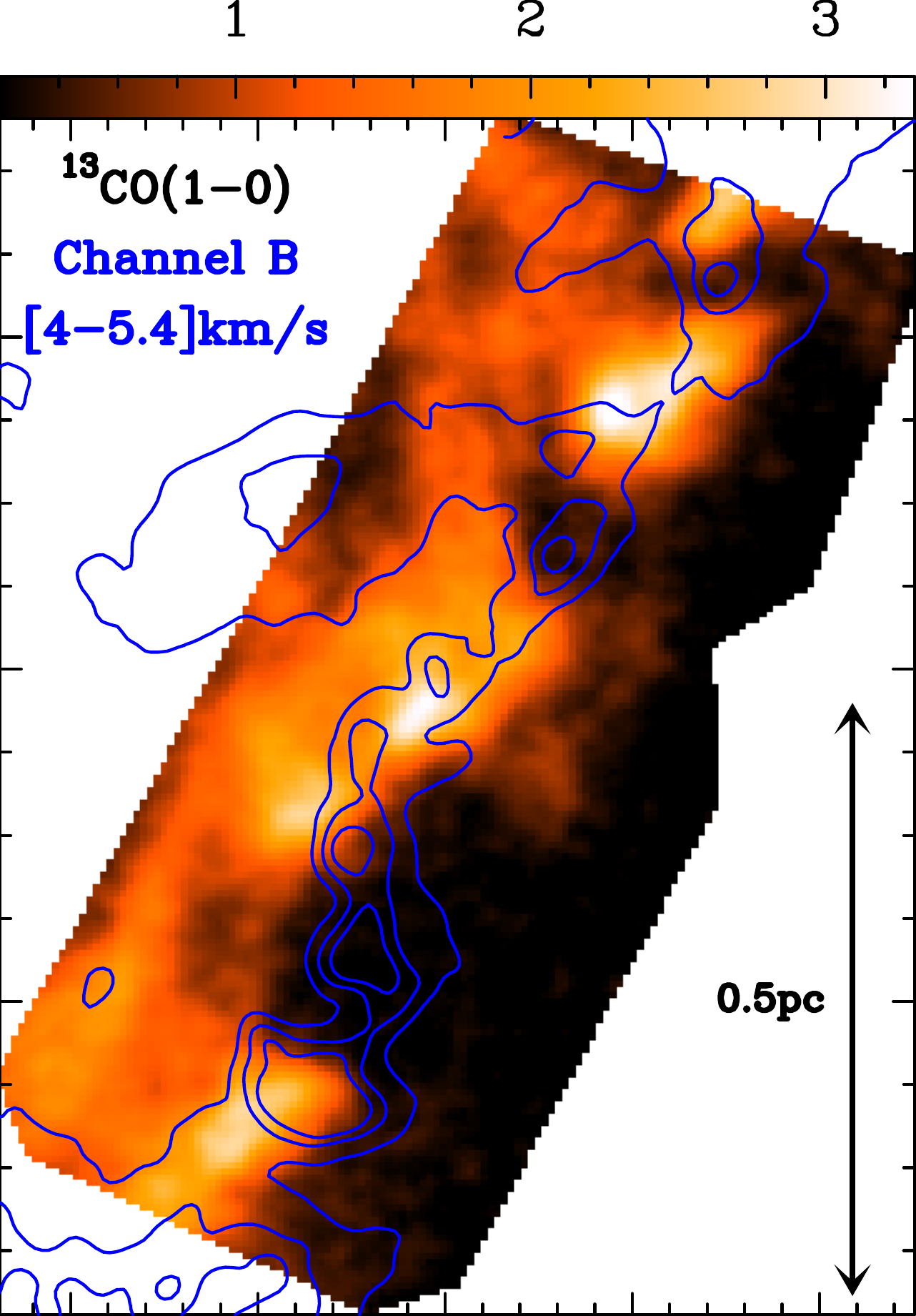} 
    \includegraphics[]{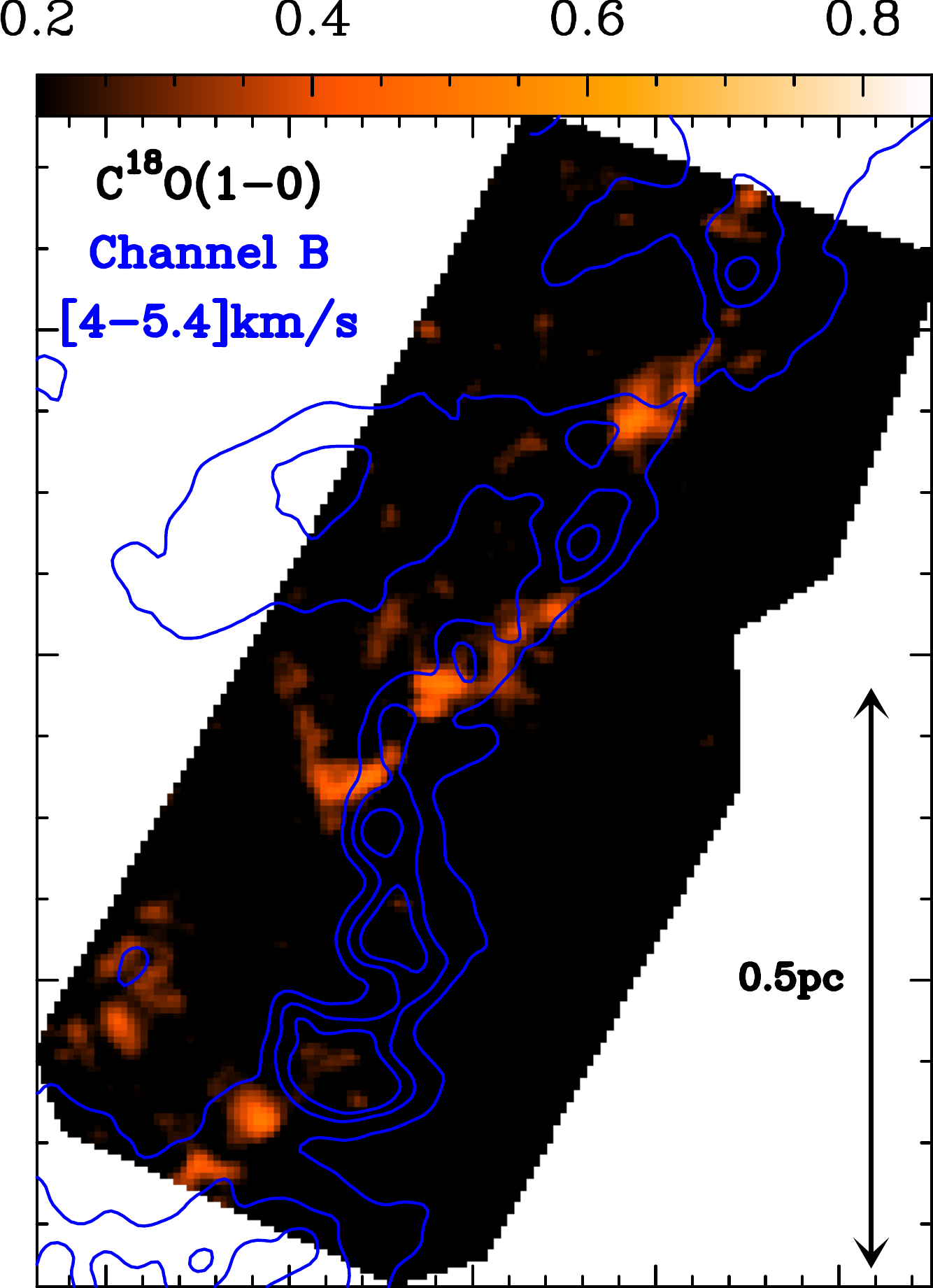} 
  }\vspace{0.5cm}
 \resizebox{15cm}{!}{  
   \includegraphics[]{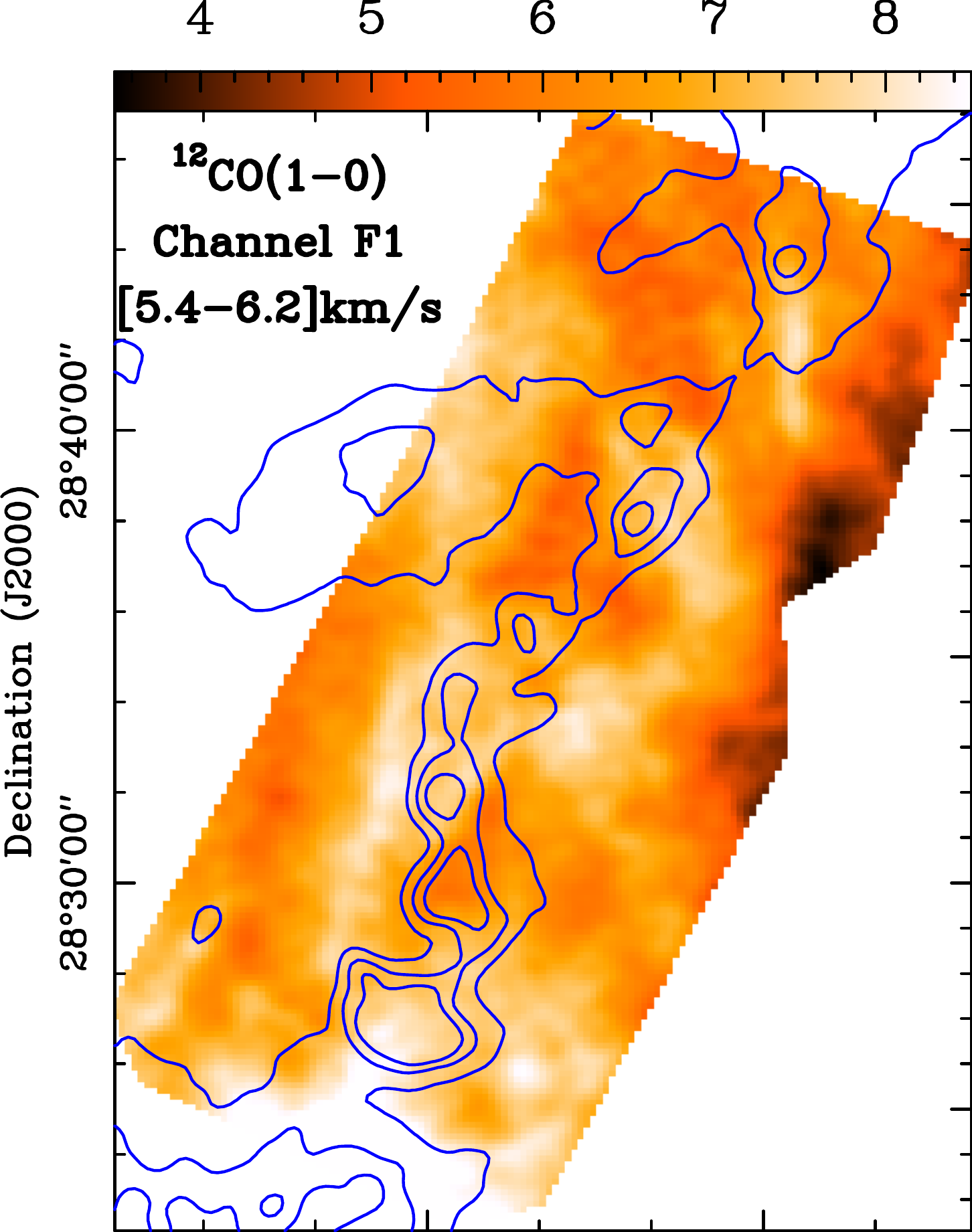} 
  \includegraphics[]{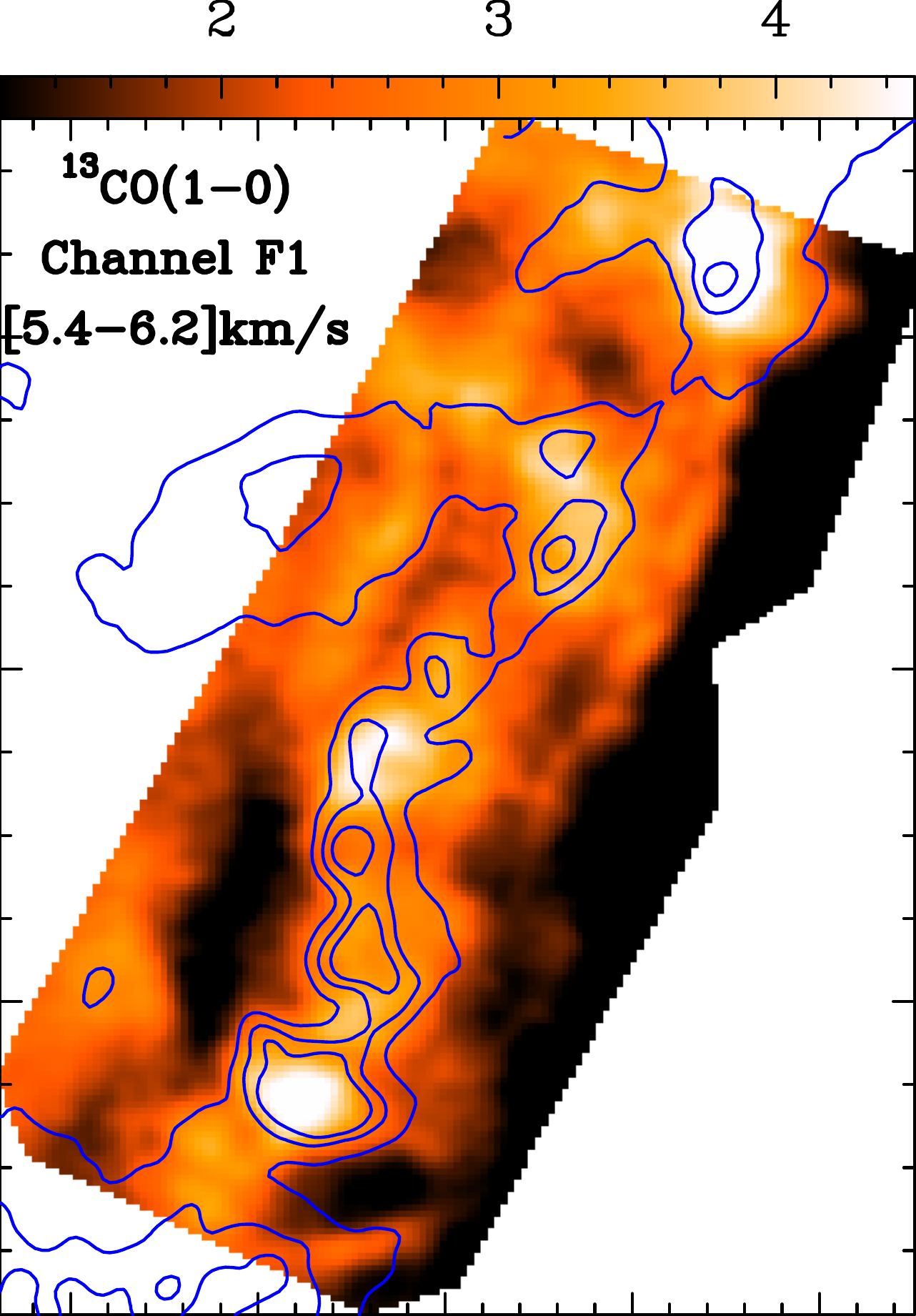} 
    \includegraphics[]{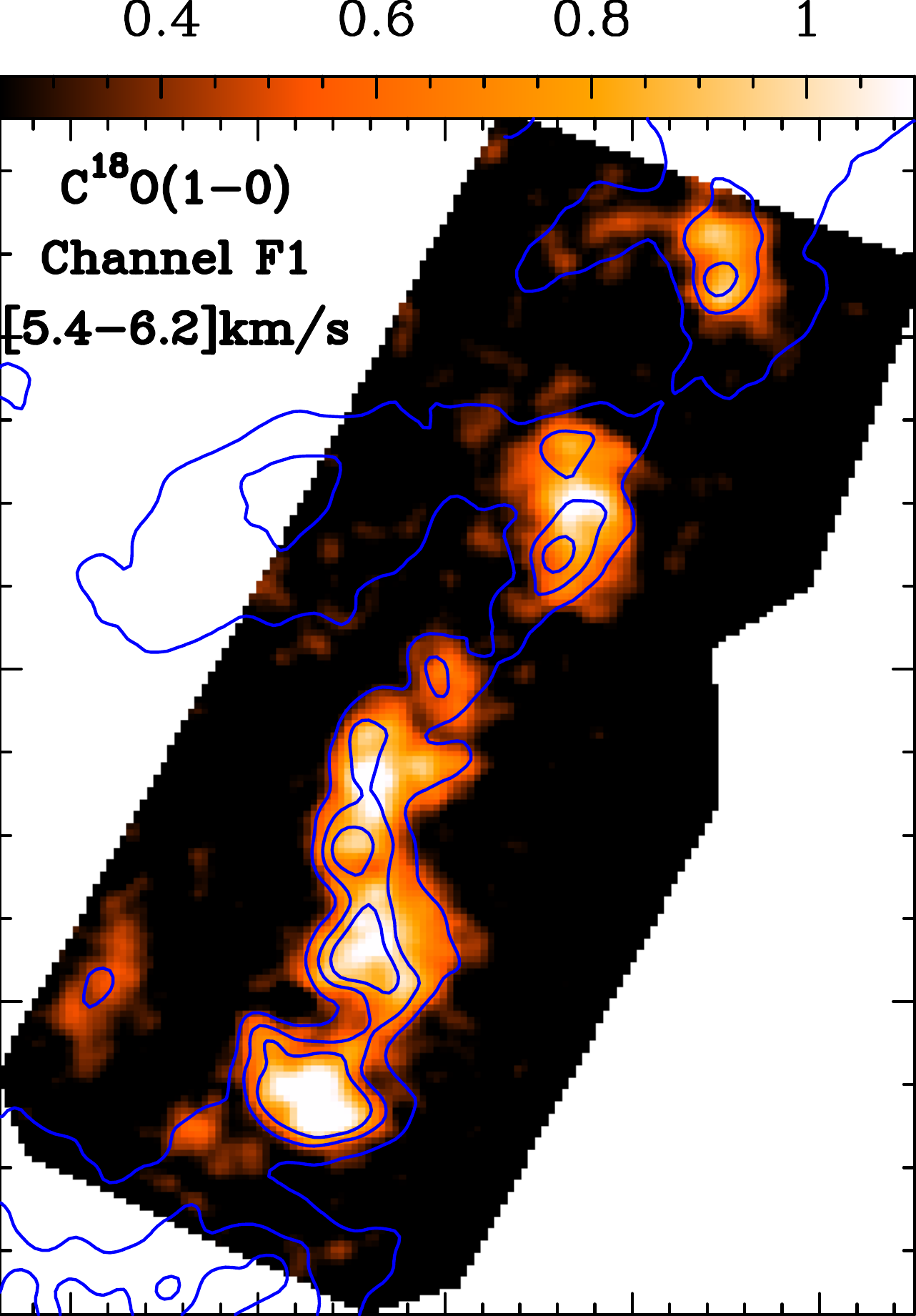} 
  }\vspace{0.5cm}
 \resizebox{15cm}{!}{
   \includegraphics[]{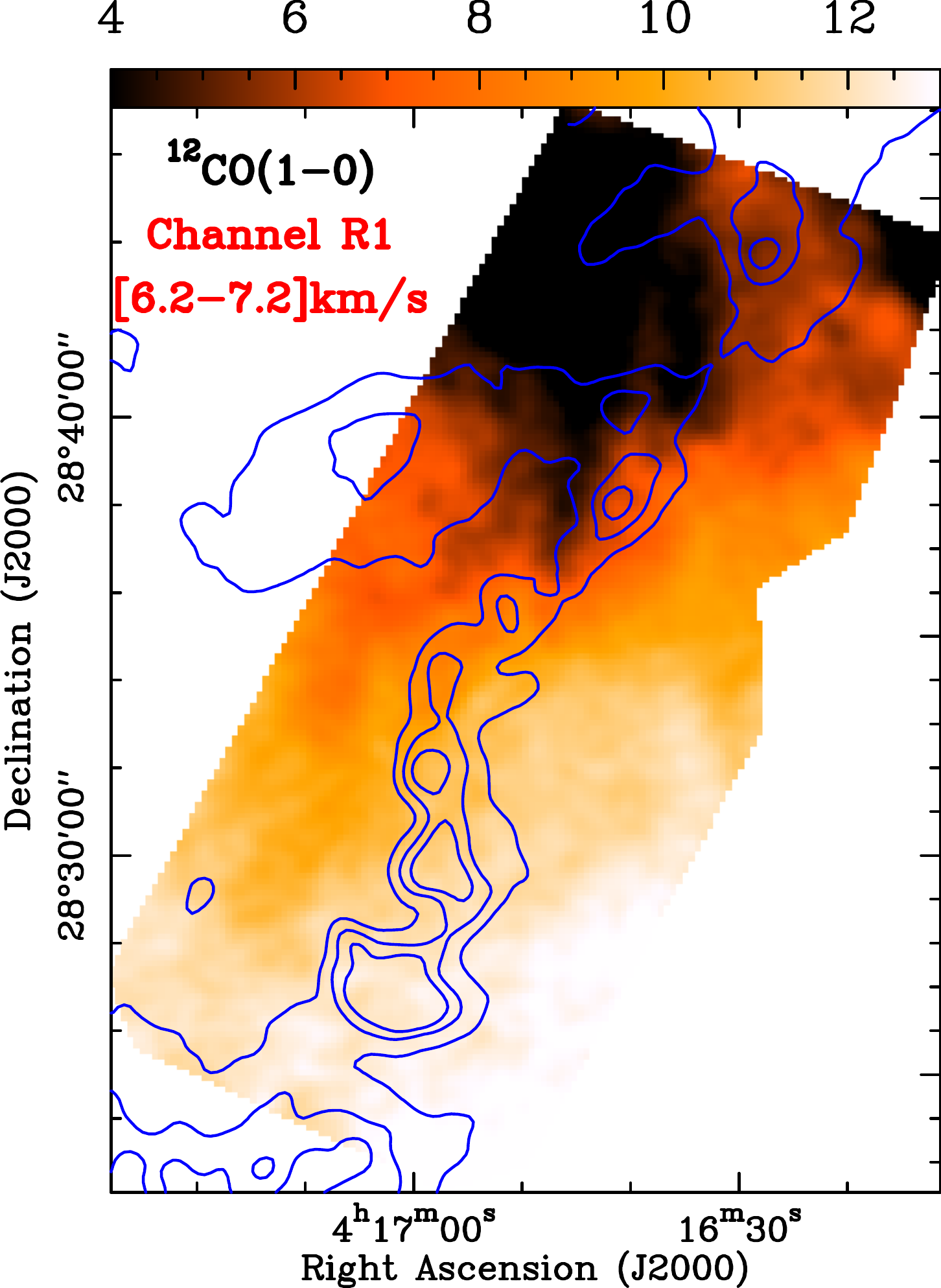} 
  \includegraphics[]{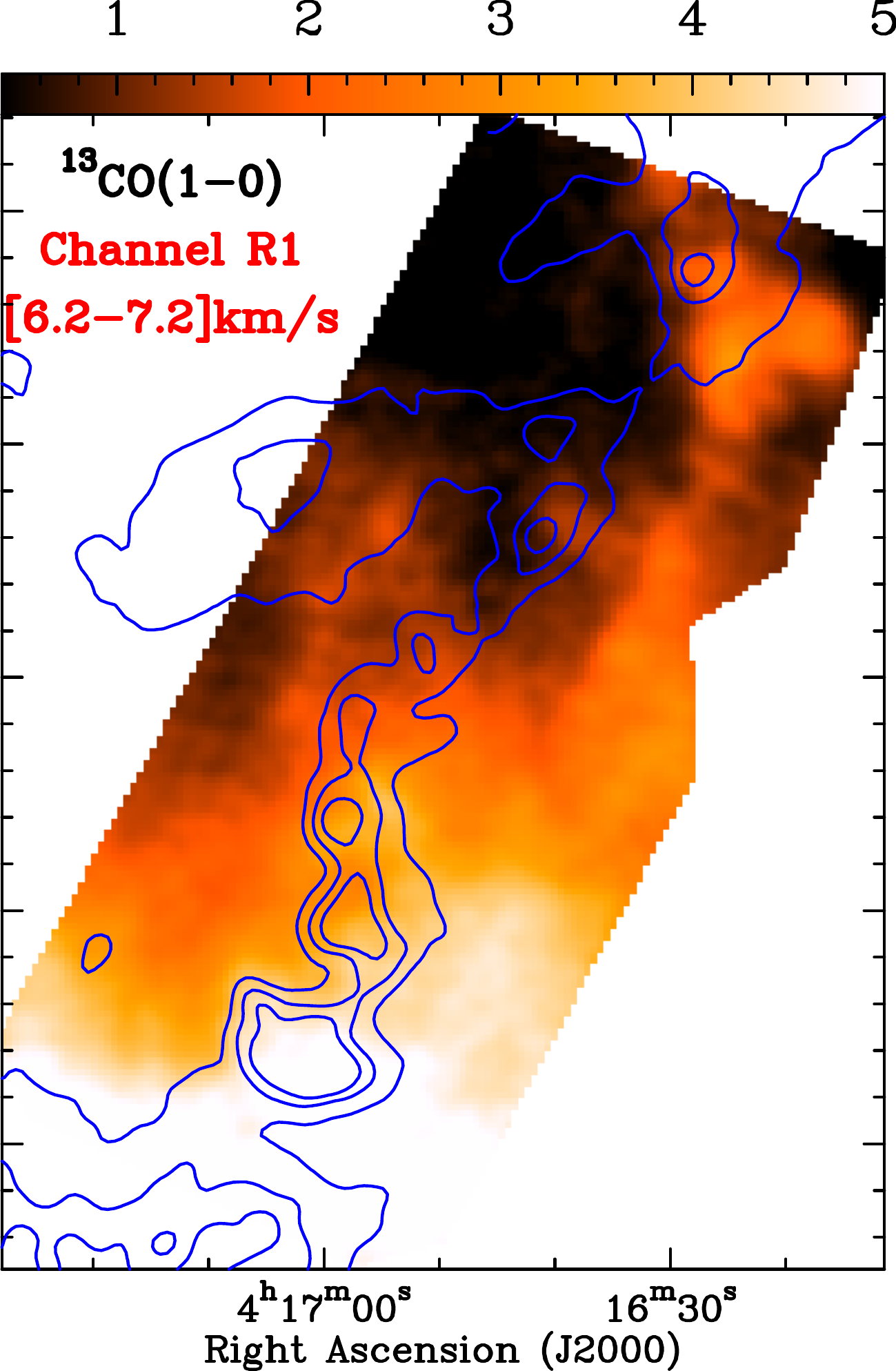} 
    \includegraphics[]{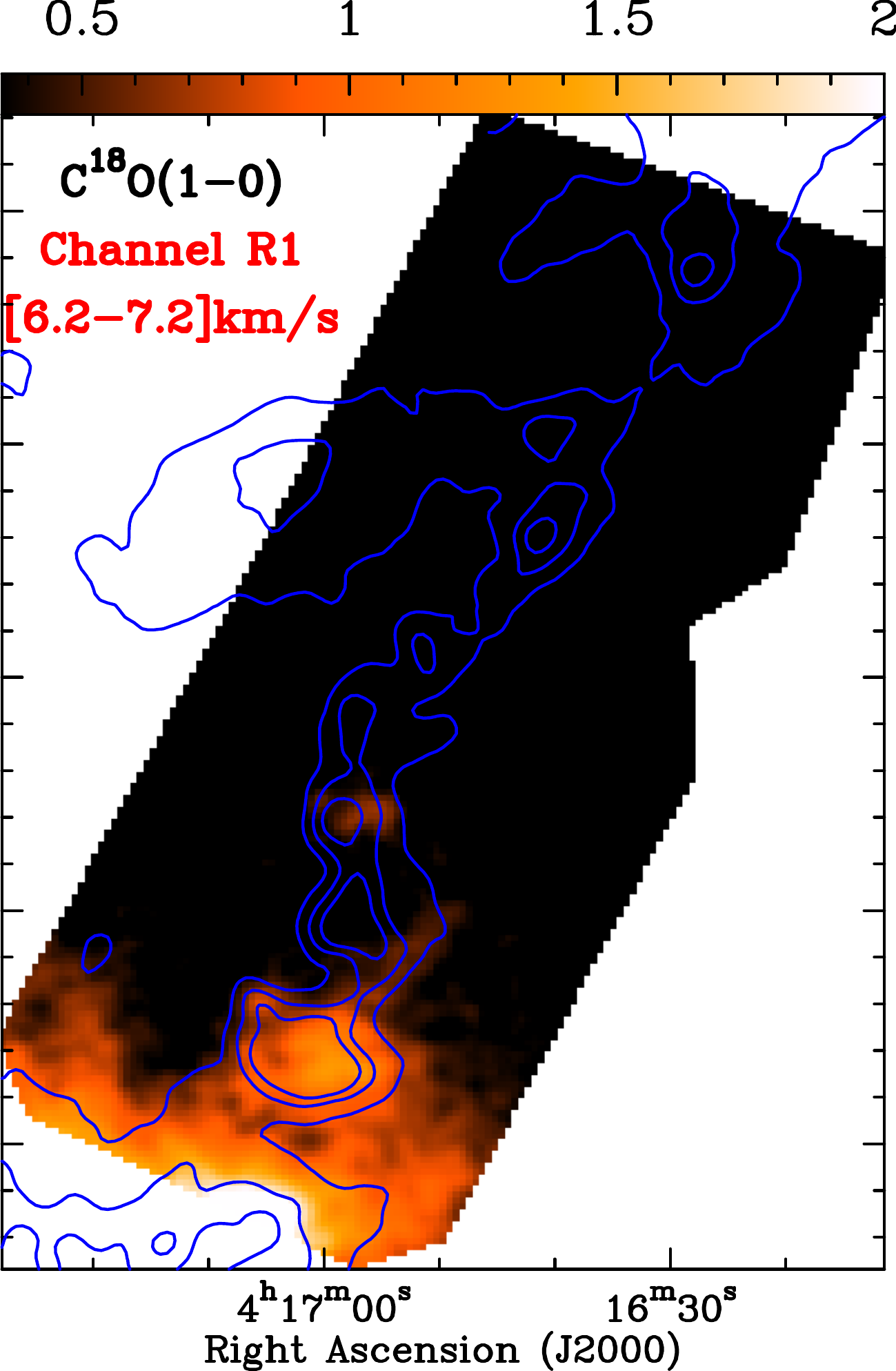} 
  }
  \end{center}
\caption{     Velocity channel maps in units of K\,\kms. 
{\rev From left to right, velocity  integrated intensity maps for $^{12}$CO$(1-0)$, $^{13}$CO$(1-0)$, and C$^{18}$O$(1-0)$ emission. 
%The maps correspond to $^{13}$CO$(1-0)$  integrated intensity apart from  the second map from the left, which shows the C$^{18}$O$(1-0)$  integrated intensity. 
From top to bottom,  channel maps   derived by integrating the intensity for the following velocity ranges: {\bf B}[4-5.4]\kms, {\bf F1}[5.4-6.2]\kms, and {\bf R1}[6.2-7.2]\kms. The velocity ranges of the channel maps are indicated on the top left of each panel. }
%with the following velocity ranges from top to bottom, 
% {\bf B}[4-5.4]\kms, {\bf F1}[5.4-6.2]\kms, and {\bf R1}[6.2-7.2]\kms, indicated on the top left of each panel. 
The spatial and spectral resolution of the maps are 28\arcsec and 0.07\,\kms, respectively.
The blue contours correspond to column densities of 4, 6, and 8\,$\times10^{21}\,\nhunit$ derived from \herschel\ data, and are the same as those in Fig.\,\ref{Herschel_map}.
{\rev Note that the range of the color scale (indicated with the bar in the top of each panel)  is different for each plot. This scale has been chosen  to  represent the dynamical range of the emission for each velocity channel map.} 
% {\bf R2}[7.2-8.1]\kms, and {\bf F2}[8.1-8.7]\kms. 
% The green  squares on the right hand side map indicate the $2'\times2'$ squares used to average the $^{12}$CO$(1-0)$ spectra plotted in Fig.\,\ref{COspectra_map} and discussed in Sect.\,\ref{Ana_12CO}. The blue oblique segment on the northeast of the field, with an orientation of $30^\circ$ (from North to East), shows the perpendicular direction to the {\bf YF} filament  used to derive the PV diagrams of Fig.\,\ref{PV_ChannelMap_Fil2}. The length of the segment correspond to 5\arcmin\ (2\parcm5 on either side of the filament crest).  
}
	        \label{CO_ChannelMaps1}
\end{figure*}

\begin{figure*}
 \begin{center} 
   \resizebox{15cm}{!}{
   \includegraphics[]{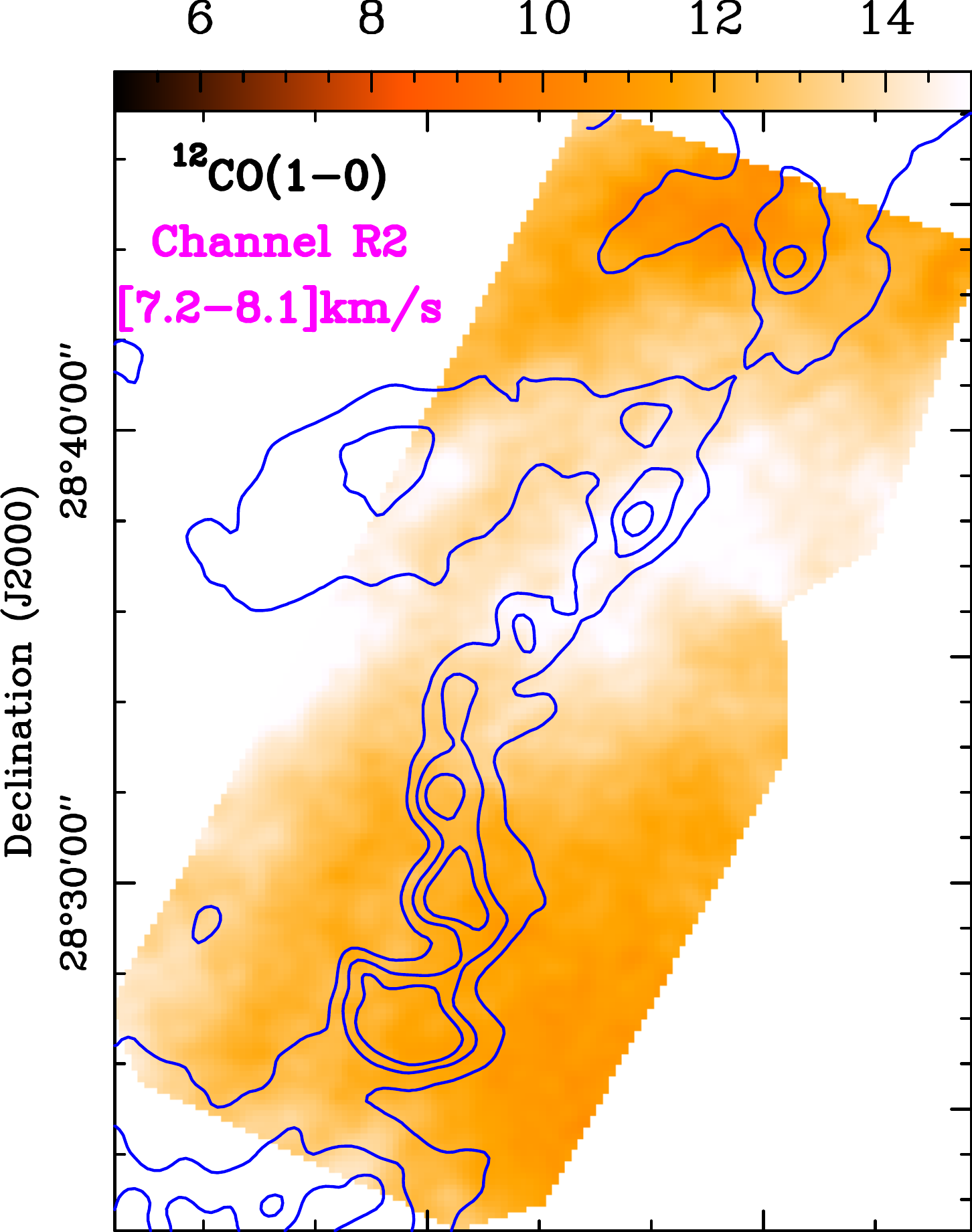} 
  \includegraphics[]{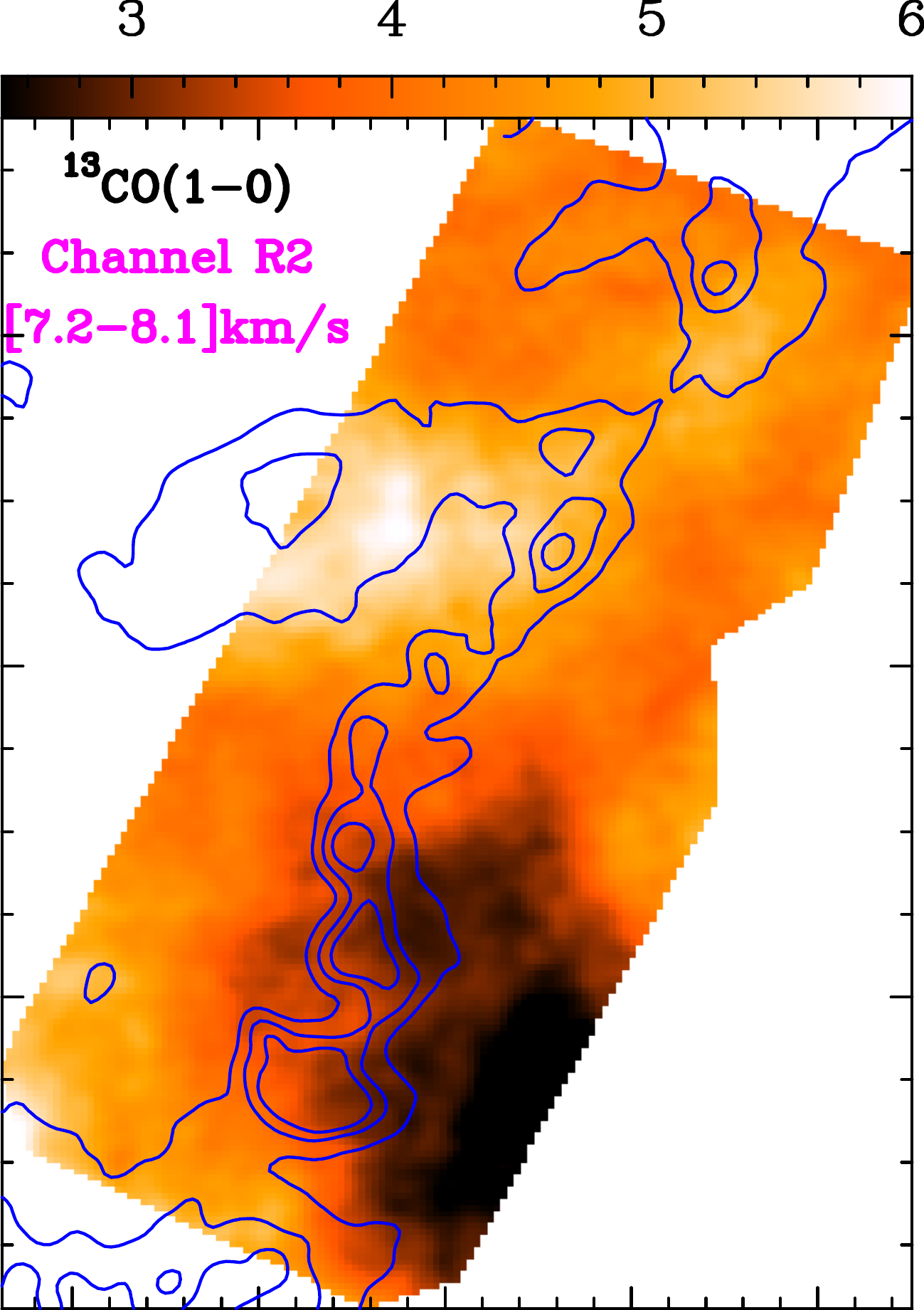} 
    \includegraphics[]{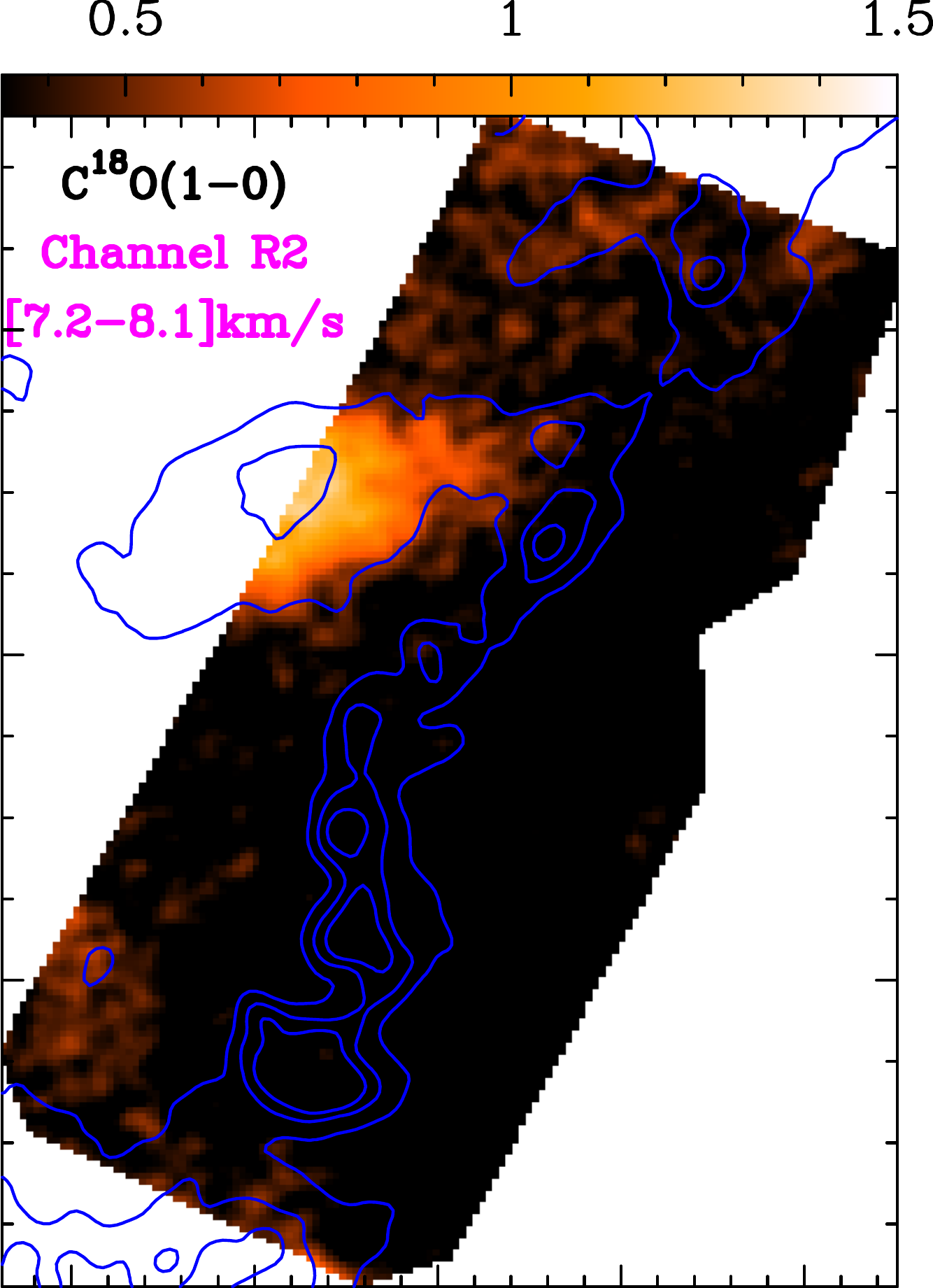} 
  }\vspace{0.5cm}
\resizebox{15cm}{!}{
   \includegraphics[]{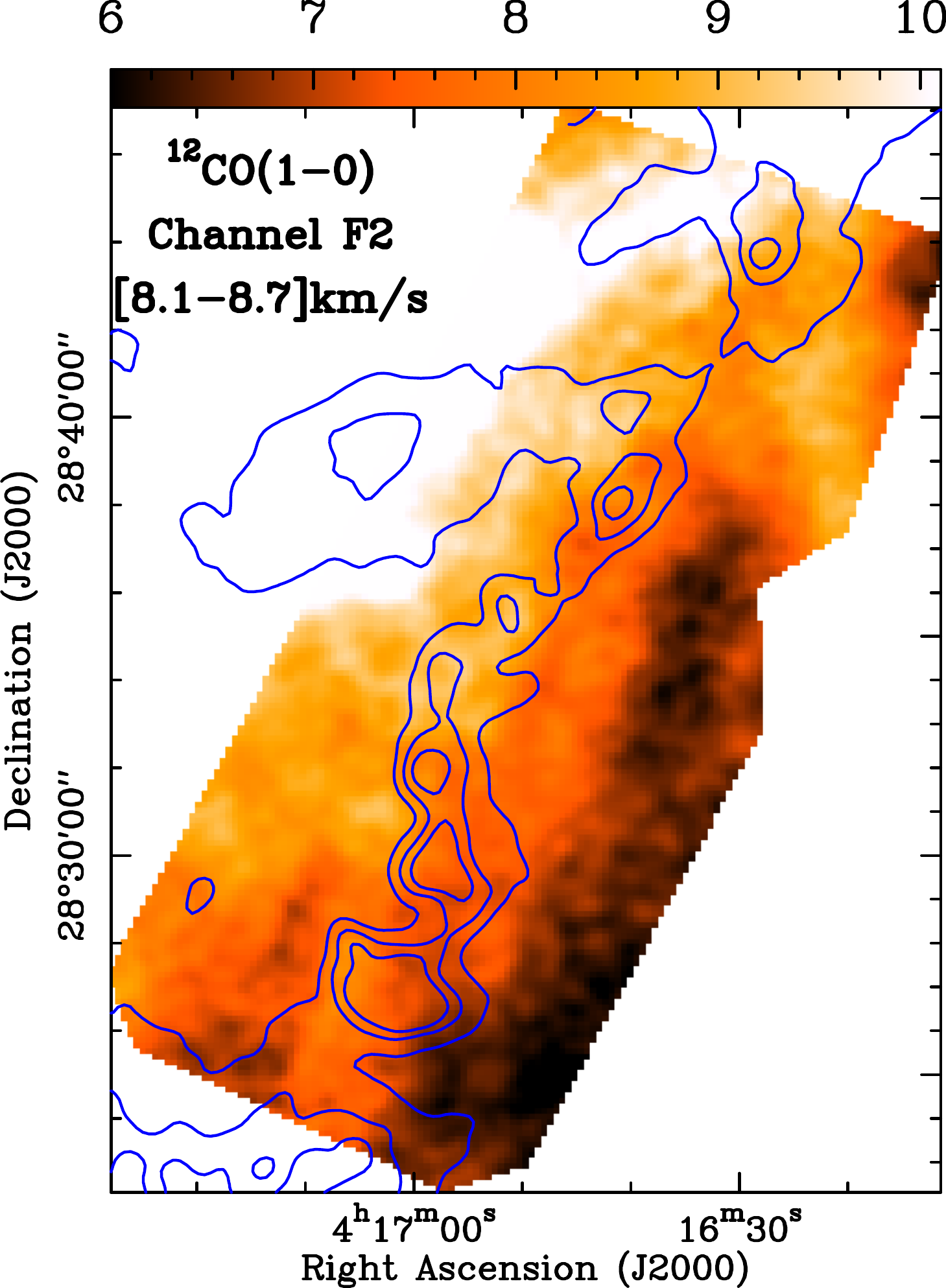} 
     \includegraphics[]{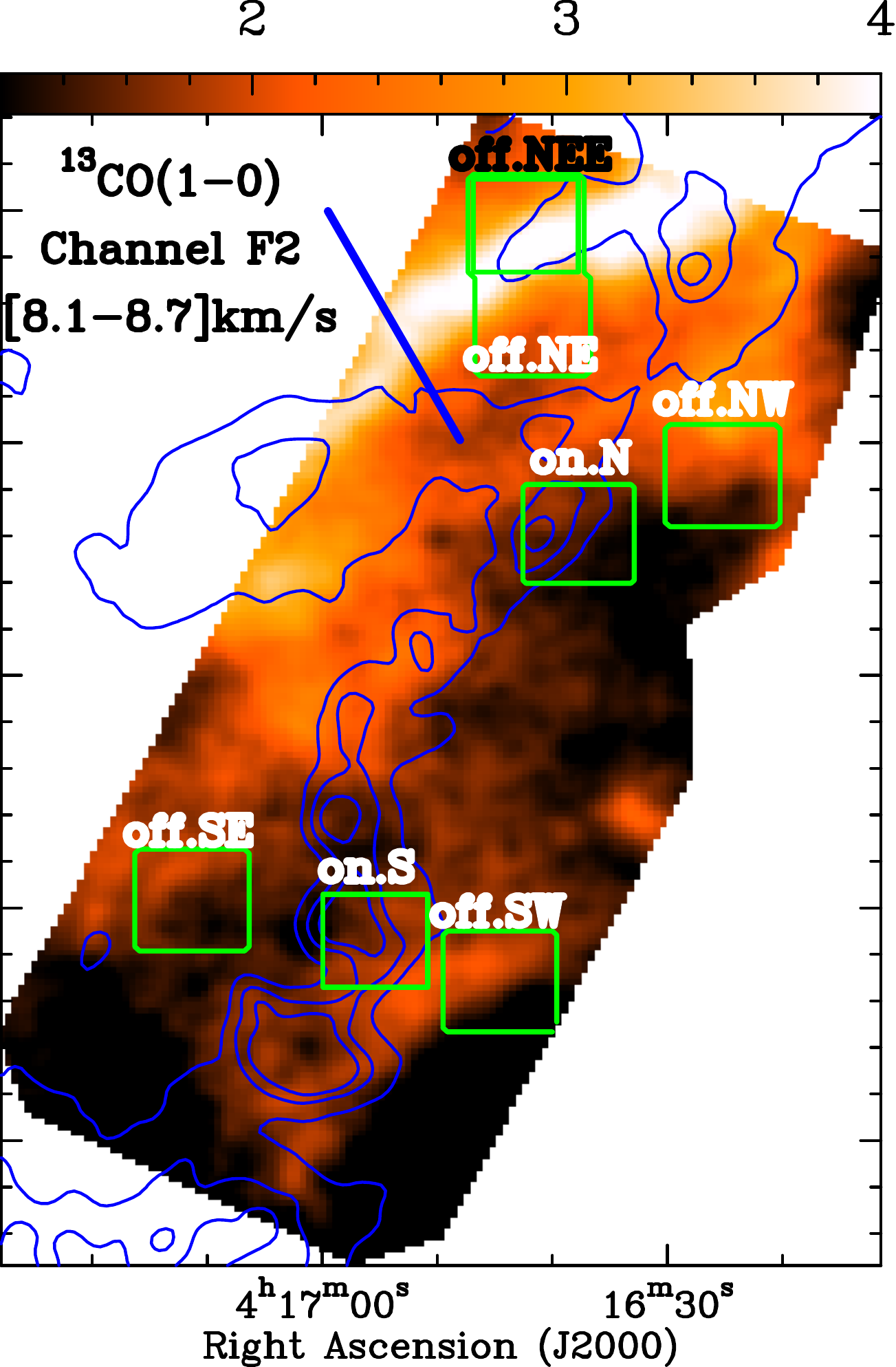}
         \includegraphics[]{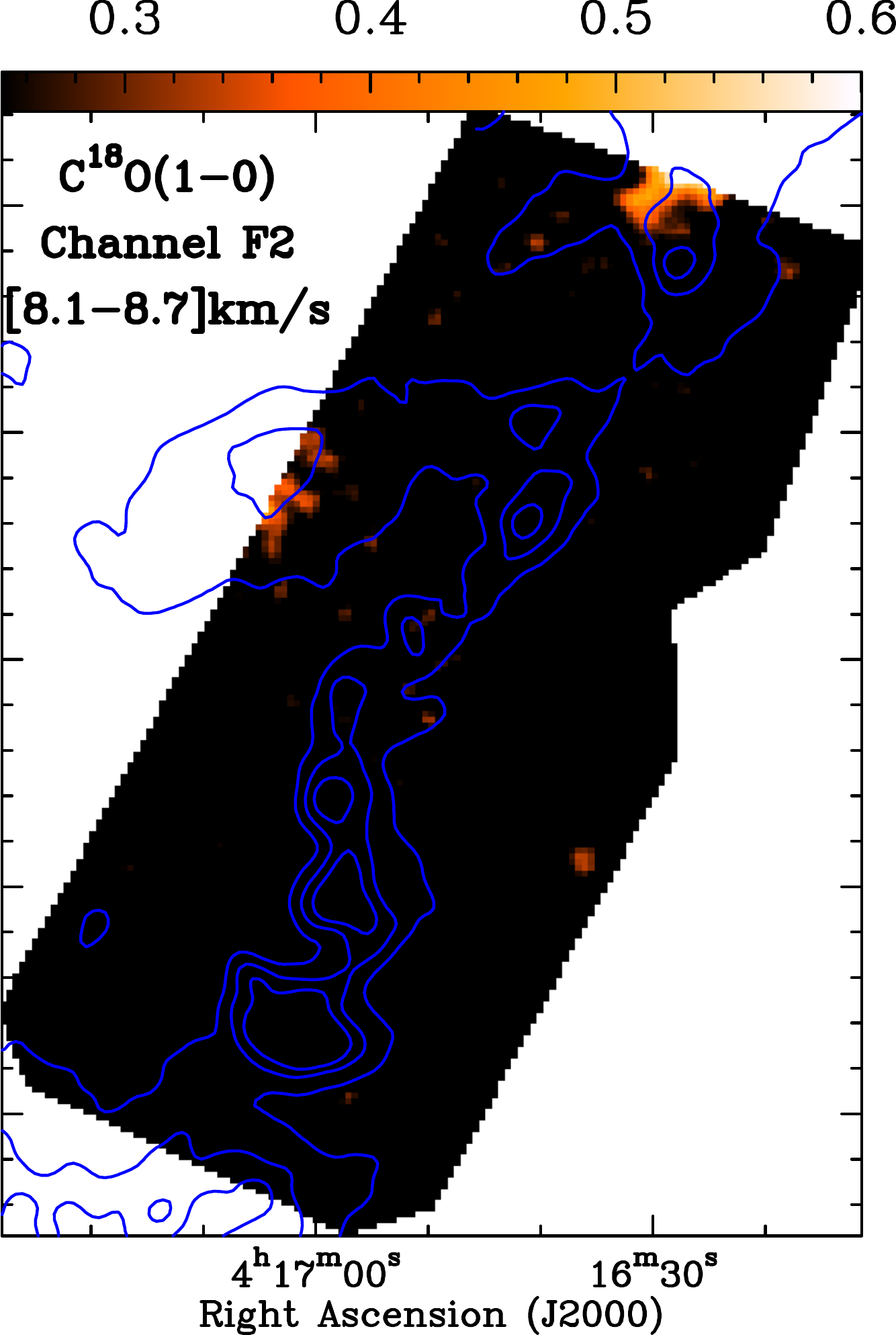} }
 \end{center}
\caption{  {\rev Same as Fig.\,\ref{CO_ChannelMaps1},    where the  channel maps are derived by integrating the intensity for  the  velocity ranges  of 
 %{\bf B}[4-5.4]\kms, {\bf F1}[5.4-6.2]\kms, {\bf R1}[6.2-7.2]\kms, 
 {\bf R2}[7.2-8.1]\kms (top) and {\bf F2}[8.1-8.7]\kms (bottom).} The green  squares on the middle-bottom panel indicate the $2'\times2'$ squares used to average the $^{12}$CO$(1-0)$ spectra plotted in Fig.\,\ref{COspectra_map} and discussed in Sect.\,\ref{Ana_12CO}. The blue oblique segment on the northeast of the field, with an orientation of $30^\circ$ (from North to East), shows the perpendicular direction to the {\bf YF} filament  used to derive the PV diagrams of Fig.\,\ref{PV_ChannelMap_Fil2}. The length of the segment correspond to 5\arcmin\ (2\parcm5 on either side of the filament crest).  
}
	        \label{CO_ChannelMaps2}
\end{figure*}

 \subsection{Position velocity diagrams and channel maps}\label{PVmaps}
 
 Figure\,\ref{CO_PV} shows position-velocity (PV) diagrams perpendicular to the main axis of the filament {\bf MF}, and averaged along its length. In practice, the cubes have first been  rotated by $25^\circ$ (corresponding to the mean orientation of {\bf MF} on the plane of the sky) from North to East. Second, the PV-cuts in the horizontal direction have been averaged  in the vertical direction along the 1\,pc length of the filament. 
 Averaging the emission increases the signal to noise ratio {\rev ($SNR$)}, which makes it possible to detect extended C$^{18}$O$(1-0)$ emission which is not strong enough to be detected  otherwise, e.g., on the channel maps ({\rev Figs.\,\ref{CO_ChannelMaps1}\,and\,\ref{CO_ChannelMaps2}).} 
 
 These PV diagrams show multiple velocity components towards the filament {\bf MF} and its surroundings. 
 From the C$^{18}$O$(1-0)$ mean PV diagram, we identify four velocity components:
  {\bf B}[4-5.4]\kms, {\bf F1}[5.4-6.2]\kms, {\bf R1}[6.2-7.2]\kms, {\bf R2}[7.2-8.1]\kms. 
  The velocities given in the brackets correspond to the velocity ranges considered for each of the four component. 
  The {\bf F1} velocity range corresponds to that of the filament {\bf MF} (cf. below Sect.\,\ref{Velstruc} and  Fig.\,\ref{CO_ChannelMaps1}).
A fifth velocity component, {\bf F2}[8.1-8.7]\kms,  is identified on the $^{13}$CO$(1-0)$ channel map as can be seen in {\rev Fig.\,\ref{CO_ChannelMaps2} (bottom middle panel)}. 
   
  \begin{figure}
 \begin{center}
  \includegraphics[width=8cm]{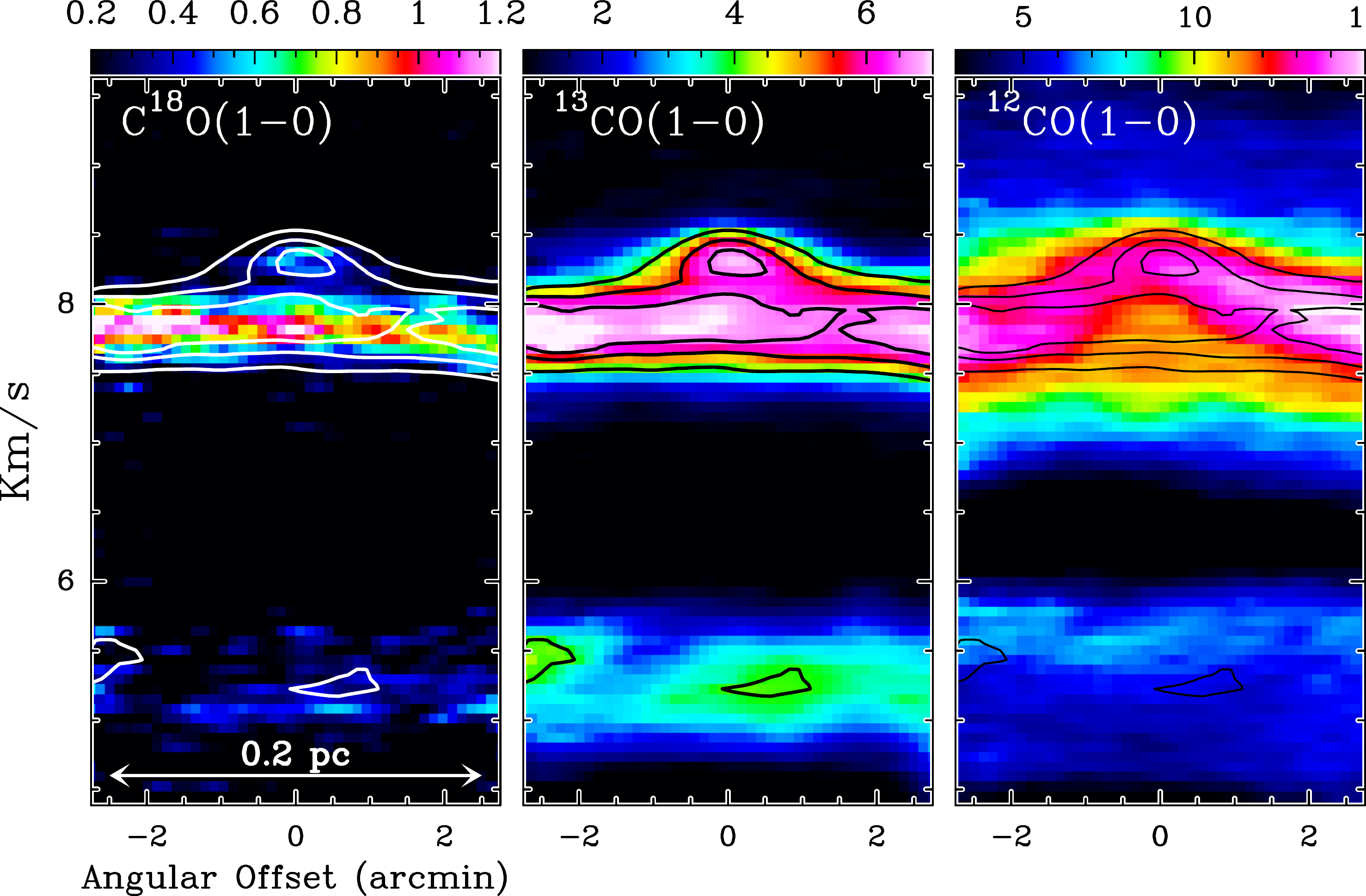} 
 \end{center}
    \caption{
	          Position velocity (PV) diagrams, {\rev in units of K($T_{\rm MB}$),} of C$^{18}$O, $^{13}$CO, and $^{12}$CO, from left to right. These PV diagrams are constructed by averaging  all the cuts taken perpendicular to the axis of {\bf YF} as traced in the  $^{13}$CO emission integrated over {\bf F2}[8.1 - 8.7]\kms\ {\rev(see bottom-middle panel of Fig.\,\ref{CO_ChannelMaps2}).}
	          %(see Fig.\,\ref{Fil_Herschel}). 
	          The zero offset position (in the x-axis) indicates the position of  {\bf YF}. The contours correspond to the $^{13}$CO emission and are the same on the three panels. 
	               } 
	   \label{PV_ChannelMap_Fil2}       
	    \end{figure}

  The analysis of the $^{12}$CO$(1-0)$  and $^{13}$CO$(1-0)$  velocity cubes indicate that the gas structures emitting at the   velocity ranges, {\bf B}, {\bf R1}, and {\bf R2}, introduced above,  have different spatial distributions towards and around {\bf MF}. 
  We name {\bf SB}, {\bf SR1}, {\bf SR2}, the structures associated to the velocity ranges  {\bf B}, {\bf R1}, {\bf R2}, respectively. {\rev In the following,  we describe the spatial distributions of these different emitting structures that can be seen on the channel maps of Figs.\,\ref{CO_ChannelMaps1}\,and\,\ref{CO_ChannelMaps2}.}
 
 {\rev As already mentioned,  the filament {\bf MF} is  traced in C$^{18}$O$(1-0)$ at the velocity range {\bf F1}, while at this same velocity range,  the $^{12}$CO emission shows an extended emission around {\bf MF}, with some decrease of the intensity towards  the filament crest (see also Sect.\,\ref{Ana_12CO}). Similarly the $^{13}$CO emission, albeit tracing {\bf MF}, has a more extended structure on both sides of the filament crest.}
  %  As can be seen on Fig.\,\ref{CO_ChannelMaps1},  
The $^{13}$CO emission of the  {\bf SB} structure, associated to the {\bf B} velocity range,  is mostly located on the eastern part of {\bf MF}, {\rev while the $^{12}$CO emission is also partly detected on the western side of  {\bf MF}}.  
  {\bf SR1} is mostly traced towards the South and North-West  of the field {\rev in both $^{12}$CO and $^{13}$CO}. {\bf SR2} covers mostly the northern part of the field  {\rev in $^{13}$CO, while $^{12}$CO emission is also seen in the South.}   At velocities between {\bf F2}[8.1-8.7]\kms, a second filament appears very bright in the North-East part of the field {\rev(see bottom-middle panel of Fig.\,\ref{CO_ChannelMaps2})}. We identify this filament as {\bf YF} (young filament, see Sect.\,\ref{youngfil} and Fig.\,\ref{Herschel_map}). 
  
{\rev The three extended structures, {\bf SB}, {\bf SR1}, and {\bf SR2}, identified from the   PV diagram of the C$^{18}$O emission, do not have sharp boundaries in space and in velocity but are interconnected spatially with   continuous velocity fields. We can  seen on Fig.\,\ref{CO_PV},  velocity gradients across  {\bf MF} from the {\bf B}  to the {\bf R1} velocity ranges, traced by both the C$^{18}$O and the $^{13}$CO emission. These velocity bridges suggest that both structures, {\bf SB} and {\bf SR1}, may be physically connected to   {\bf MF} with partly mixed  velocities.   
On the other hand, {\bf SR2} may not be presently directly connected to  {\bf MF} neither in velocity and probably nor in space. 
%The PV diagram of $^{12}$CO emission  shows an extended structure at the velocity range {\bf F1}, with a larger intensity than the $^{12}$CO emission at the velocity range {\bf B} (Fig.\,\ref{CO_PV}-right). 
}

The PV diagrams perpendicular to {\bf YF} (Fig.\,\ref{PV_ChannelMap_Fil2}) show two extended  structures, one at velocities $\sim5$\kms\ and another at around $\sim8$\kms. These two velocity components correspond to the channels {\bf B}  and {\bf R2}, respectively (see Fig.\,\ref{PV_ChannelMap_Fil2}). The filament {\bf YF} is identified  on these PV diagrams in both C$^{18}$O$(1-0)$ and $^{13}$CO$(1-0)$, as a compact structure at a velocity $\sim8.3$\kms. On the $^{12}$CO$(1-0)$ PV diagram an extended structure is detected at velocities between $\sim7.5$ and 8.5\kms, with a  bent around the position of {\bf YF} (Fig.\,\ref{PV_ChannelMap_Fil2}, see Sect.\,\ref{youngfil}).

 	       \begin{figure}
 \begin{center}
  \includegraphics[width=8cm]{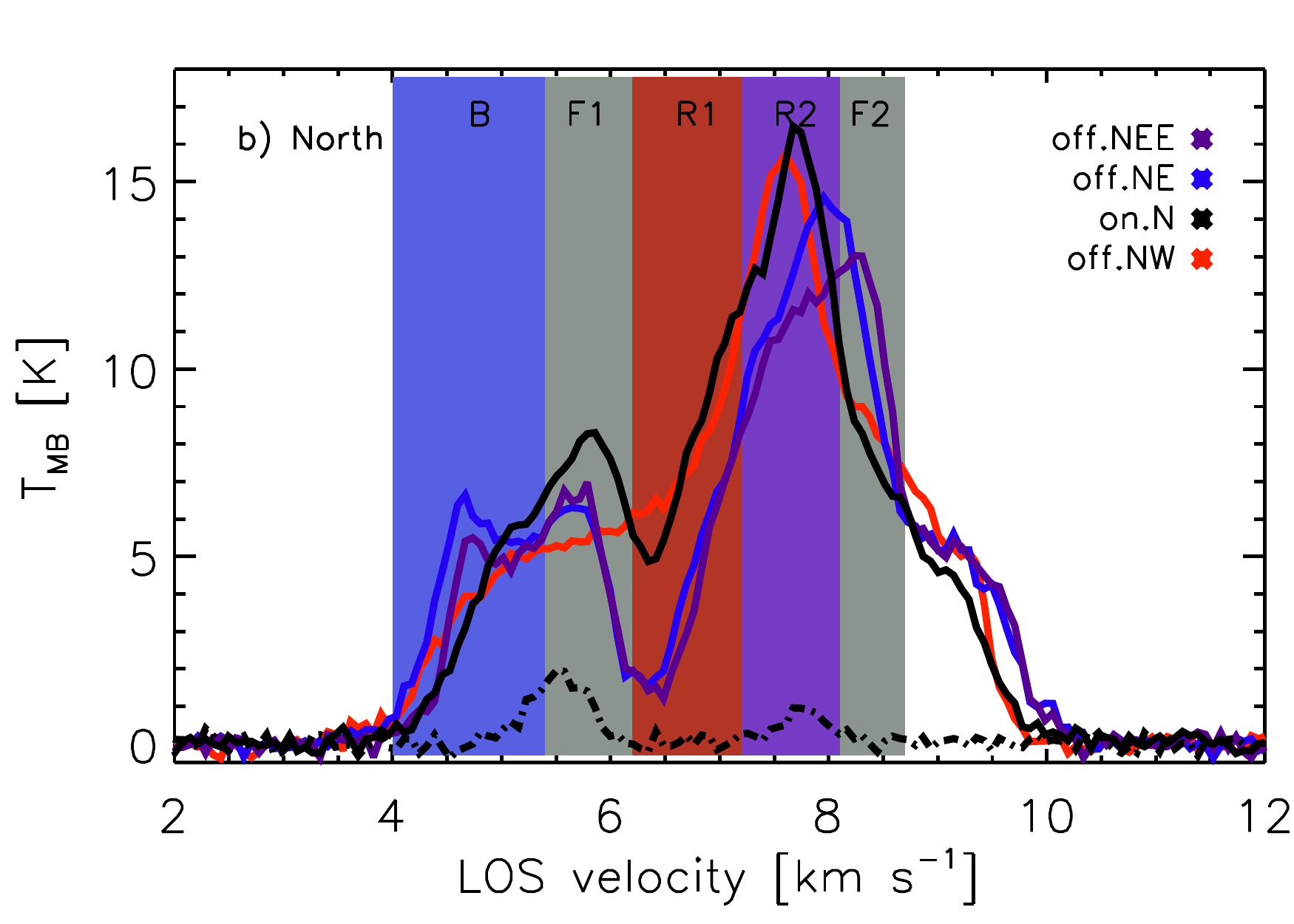} 
  \includegraphics[width=8cm]{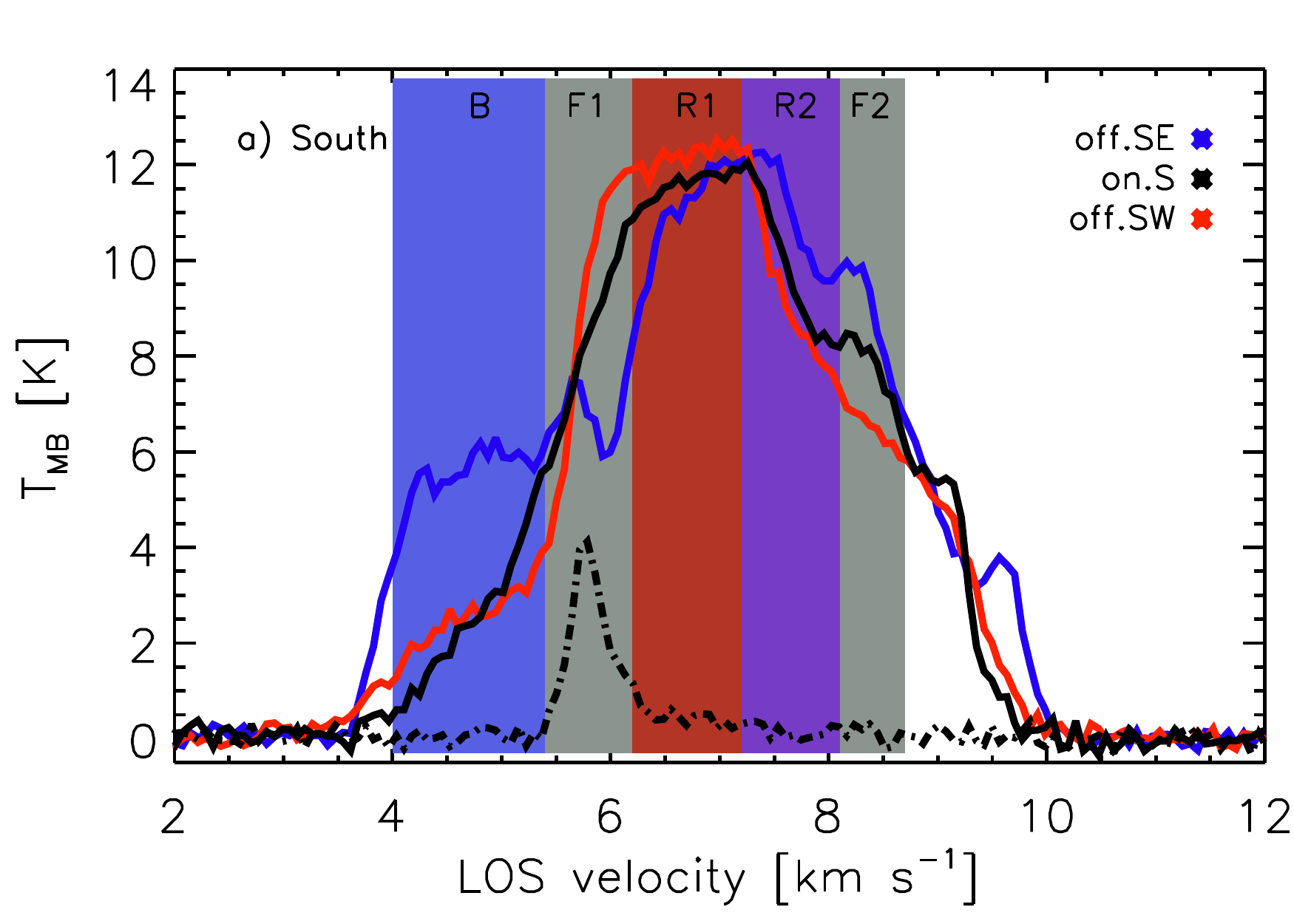} 
 \end{center}
\caption{ The solid line spectra correspond to the $^{12}$CO$(1-0)$ emission integrated within $2'\times2'$ boxes on and off the filament, towards the North (a) and South (b) of the observed region. The color coding is indicated in the top right of the panels.The dotted spectra show integrated  C$^{18}$O$(1-0)$ emission within the same  $2'\times2'$ boxes observed on the filament.
 The Five colored polygons correspond to the 5 channels defined in Sect.\,\ref{Velstruc}:  
 {\bf B}[4-5.4]\kms, {\bf F1}[5.4-6.2]\kms, {\bf R1}[6.2-7.2]\kms, {\bf R2}[7.2-8.1]\kms, {\bf F2}[8.1-8.7]\kms.  
	               }
	   \label{COspectra_map}       
	    \end{figure}
	    
	\subsection{Analysis of $^{12}$CO$(1-0)$ spectra}\label{Ana_12CO}  

The velocity channel maps (Figs.\,\ref{CO_ChannelMaps1}\,and\,\ref{CO_ChannelMaps2}) and the PV diagrams (Fig.\,\ref{CO_PV}) show the presence of extended structures (with respect to the compact, elongated shape of the main filament {\bf MF}). These extended structures have different velocity components at redshifted and blueshifted LOS velocities with respect to the mean velocity detected in C$^{18}$O$(1-0)$  along {\bf MF}. 

To get a hint on the relative position of the different extended structures surrounding {\bf MF}, 
  we analyse the $^{12}$CO$(1-0)$ spectra. To do so, we compare the optically thick $^{12}$CO$(1-0)$ spectra at different positions ``on" and ``off" {\bf MF}  searching for signatures of absorption that may be used as an indication of the relative position of the different emitting structures along the LOS. 
Figure\,\ref{COspectra_map} shows $^{12}$CO$(1-0)$ spectra integrated in $2'\times2'$ squares centered ``on"  {\bf MF} in the North and South, as well as ``off" {\bf MF} in the West and East sides of the  filament axis. 
The $^{12}$CO$(1-0)$  ``on"  spectra in the {\bf B} velocity range, both in the North and in the South have lower intensity  than the ``off" spectra in the same velocity range (comparing the blue and black spectra of Fig.\,\ref{COspectra_map} in the  {\bf B} velocity range). This decrease in the $^{12}$CO$(1-0)$ intensity may be {\rev interpreted as} due to   
absorption of the optically thick emission {\rev by the gas along the filament at velocities in the {\bf B} range and colder than the gas of the {\bf SB} structure}. 
These  absorption features may suggest that, along the LOS, the {\bf SB}   structure  is  located (at least partly) in the back of  the different structures observed towards the field,  including {\bf MF}. On the other hand, such variations of the $^{12}$CO$(1-0)$ intensity ``on"  and ``off"   are not observed neither at  the {\bf R1}  nor at the {\bf R2} velocity ranges (comparing the red and black spectra of  Fig.\,\ref{COspectra_map} in the {\bf R1} and {\bf R2} velocity ranges),  suggesting that the {\bf SR1} and {\bf SR2}  structures may be located mostly in the front of {\bf SB} and {\bf MF} along the LOS. The $^{12}$CO$(1-0)$ spectra towards the North-East side shows no  absorption features  ``on"  {\bf YF} (comparing the red and purple spectra of Fig.\,\ref{COspectra_map}a in the {\bf F2} velocity range) suggesting that {\bf YF} is located in front of the other emitting structures along the LOS.  {\rev We note that such intensity variations in the $^{12}$CO emission may also result from structures with different  spatial distribution around  {\bf MF}.  Complementary observations of, e.g., other transitions, and radiative transfert modeling, may be needed to establish the physical reason behind the variations of the $^{12}$CO emission.} %We  discuss this aspect further in Sect.\,\ref{filcloud}.}
In Appendix\,\ref{AppSpectra}, we present {\rev $^{12}$CO, $^{13}$CO, and C$^{18}$O  spectra  averaged in $2'\times2'$ squares  towards the whole observed region (Fig.\,\ref{COspectra_map}).}

 \subsection{Velocity structure along and around {\bf MF}}\label{Velstruc}

 \begin{figure}
 \begin{center}
  \includegraphics[width=8cm]{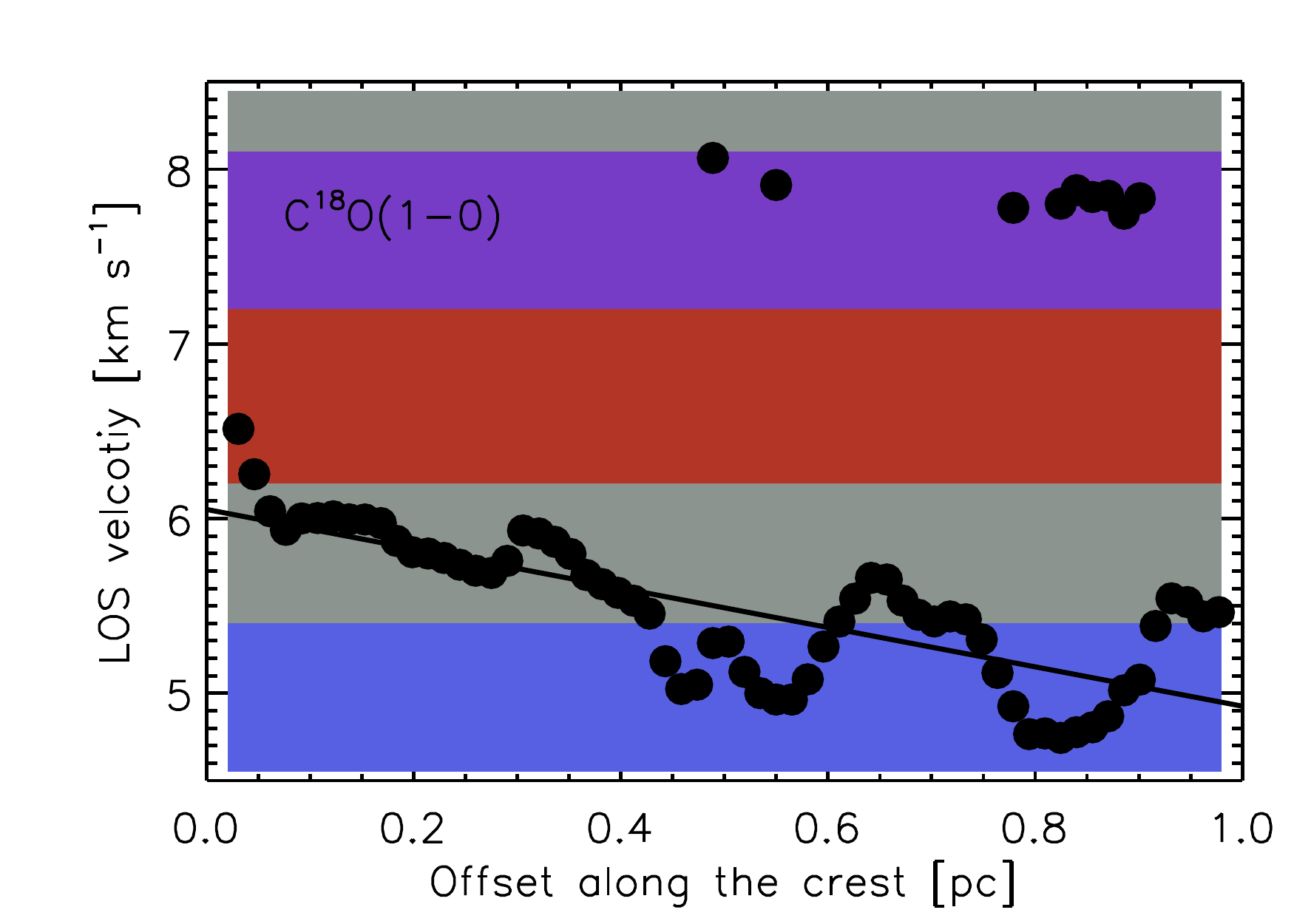} 
 \end{center}
\caption{Line of sight  velocity derived from C$^{18}$O$(1-0)$ spectra along the crest of {\bf MF} from South to North (from $>0$ to $\sim1$\,pc). {\rev The filament crest is shown on Fig.\,\ref{Herschel_map}-right.} Each dot corresponds to averaged  spectra within one beam of 28\arcsec. 
The oblique line shows the best linear fit to the velocity structure as $\langle v_{\rm LOS}\rangle^{\rm grad}=(6.05\pm0.07$\kms\,pc$^{-1})\,v_{\rm LOS}+(-1.13\pm0.12$\kms).
The dots at velocity of $\sim8$\kms are observed towards the filament, but are linked to a different structure. 
 The Five colored polygons correspond to the 5 channels defined in Sect.\,\ref{Velstruc}. From bottom to top:  
 {\bf B}[4-5.4]\kms, {\bf F1}[5.4-6.2]\kms, {\bf R1}[6.2-7.2]\kms, {\bf R2}[7.2-8.1]\kms, {\bf F2}[8.1-8.7]\kms.  
}
	        \label{vel_alongfil}
\end{figure}

 The analysis of the $^{12}$CO$(1-0)$  and $^{13}$CO$(1-0)$  velocity cubes indicate several velocity components in the surroundings of the filament {\bf MF}. These various velocity structures have different spatial distributions in the observed field (Figs.\,\ref{CO_ChannelMaps1}\,and\,\ref{CO_ChannelMaps2}). 
 
 The filament {\bf MF} is detected in C$^{18}$O$(1-0)$ as a velocity coherent structure with mostly a single velocity component all along the crest (Fig.\,\ref{vel_alongfil}).
 Towards the North section of {\bf MF} the C$^{18}$O$(1-0)$ spectra show a second velocity component at $\sim 8$\,\kms. 
  The structure emitting at this systemic velocity is probably  not  connected neither spatially nor in velocity to {\bf MF} (cf. Fig.\,\ref{CO_PV}). 
 
\begin{figure}
 \begin{center}
 \hspace{-0.5cm}
  \includegraphics[width=8.5cm]{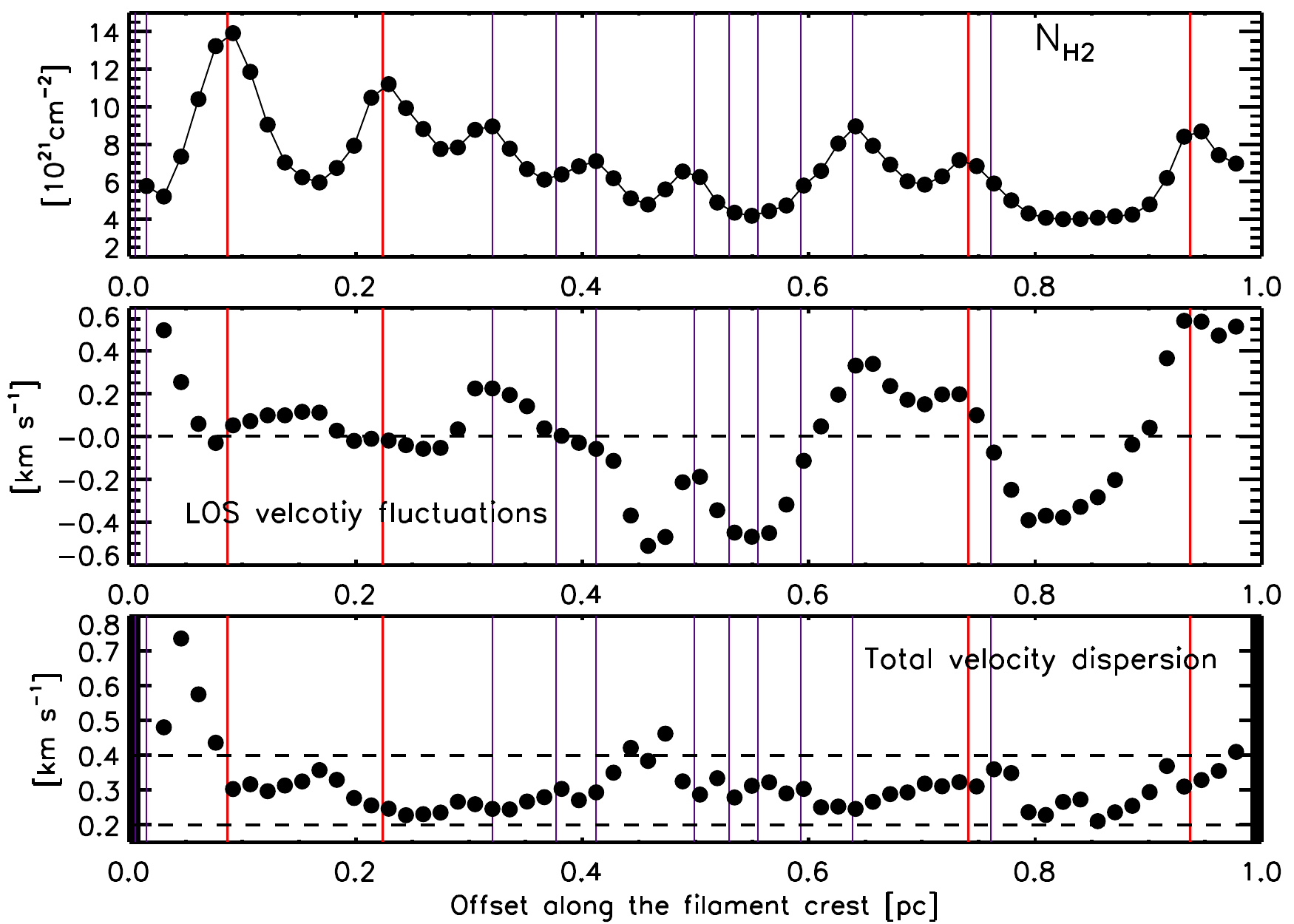} 
 \end{center}
\caption{Properties along the crest of {\bf MF} from South to North. Each dot corresponds to averaged  values within one beam of 28\arcsec. On all three plots, the vertical red and purple lines indicate the positions of the prestellar and starless cores, respectively,  identified with  \herschel\ along the filament crest  (cf. Fig.\,\ref{Herschel_map}). 
{\bf Top:} Column density derived from \herschel\ data. 
{\bf middle:} Line of sight  velocity fluctuations $\delta v_{\rm LOS}$ derived from C$^{18}$O$(1-0)$ spectra along the crest, where $\delta v_{\rm LOS}=v_{\rm LOS}-\langle v_{\rm LOS}\rangle^{\rm grad}$ (cf. Fig.\,\ref{vel_alongfil}).  The horizontal dashed line indicates the zero value. 
{\bf Bottom:} Total velocity dispersion $\sigma_{\rm tot}$ along the crest derived from C$^{18}$O$(1-0)$ spectra. The horizontal dashed lines indicate the values of the sound speed $c_{\rm s}=0.2$\kms  and $2c_{\rm s}$, for $T_{\rm gas}=10$\,K. 
}
	        \label{alongfil_3panel}
\end{figure}

  Figure\,\ref{vel_alongfil} shows a large scale velocity gradient of $\sim1$\,\kms\,pc$^{-1}$ along the filament.   
   {\rev  In the southern part, for $0<x_{\rm crest}<0.4$\,pc, where $x_{\rm crest}$ is the position along the filament crest, {\bf MF} has a velocity in between the velocities of the extended structures {\bf SB} and {\bf SR1}, detected at velocity ranges blueshifted and redshifted, respectively, with respect to  that of {\bf MF}.} 
 Towards the North, {\rev for $x_{\rm crest}>0.4$, } {\bf MF} has a velocity  compatible with the velocity of the extended structure {\bf SB} detected at velocity ranges {\bf B}, corresponding to  blueshifted velocities with respect to  that of {\bf MF}. 
  %  The observed velocity gradient along the filament, connects the filament velocity  from the South to the velocity of the extended structure {\bf SR1} observed at velocity ranges of  {\bf R1},  redshifted with respect to mean velocity observed in C$^{18}$O$(1-0)$ along  {\bf MF}. 
 On top of this large scale velocity gradient, along the southern part of {\bf MF}  we can see small scale velocity fluctuations with an amplitude of about 0.13\,\kms\ %$\delta v_{\rm LOS}=0.13$\,\kms\ (standard deviation  of $v_{\rm LOS}$) 
 for $0<x_{\rm crest}<0.4$\,pc. % where $x_{\rm crest}$ is the position along the filament crest. 
 Along the northern part of {\bf MF} the velocity fluctuations have  larger amplitudes of about %$\delta v_{\rm LOS}=
 $0.34$\,\kms\ for $x_{\rm crest}>0.4$\,pc, as can been seen in the middle panel of Fig.\,\ref{alongfil_3panel}.

\begin{table*}[!h] 
  \tbl{Median values of optical depth, column density, and line of sight length for the different structures identified in velocity space. {\rev The  median values are computed over the observed field, for each velocity range, and for a signal to noise ratio $SNR>5$. \\}}  
 { \begin{tabular}{|c|c|cccc|cccc|c|}
      \hline
     \hline
    Structure & Velocity range &  $\tau_{^{13}\rm{CO}}$&$FWHM_{^{13}\rm{CO}}$&$^{13}$\nhh&$^{13}l_{\rm{LOS}}$&$\tau_{\rm{C}^{18}\rm{O}}$&$FWHM_{\rm{C}^{18}\rm{O}}$&$^{18}\nhh$ &$^{18}l_{\rm{LOS}}$ \\% &$^\herschel\nhh$ \\
      & [\kms]  & &[\kms]&[$10^{21}$\nhunit] &[pc]&&[\kms]&[$10^{21}$\nhunit] &[pc] \\% &[$10^{21}$\nhunit] \\  
     (1)&(2) & (3)&(4)&(5)&(6)&(7)&(8)&(9)&(10) \\ %&(11)\\  
      \hline
      % from mom2 and peak intensity derived from channel maps
  SB  &\,\,\,\,\, B[$4-5.4$] & 0.80&     0.68&      1.73&     0.28&     0.15&     0.70&      3.82&     0.25\\
     MF &F1[$5.4-6.2$]&0.75&     0.51&      1.31&     0.21&     0.20&     0.44&      3.06&     0.20\\
       SR1 &R1[$6.2-7.2$]&0.79&     0.68&      1.76&     0.29&     0.21&     0.58&      4.35&     0.28\\
        SR2 & R2[$7.2-8.1$]&1.09&     0.66&      2.45&     0.40&     0.16&     0.47&      2.80&     0.18\\
  YF &F2[$8.1-8.7$]&0.67&     0.42&     0.97&     0.16&     0.14&     0.40&      1.98&     0.13\\
        \hline
     All$^*$   &&0.79&     0.66&      1.73&     0.28&     0.16&     0.47&      3.06&     0.20\\
           \hline
    \end{tabular}}
    \label{tab:coldens_tau}
\begin{tabnote}
{\bf Col. 1}: Name of the structure identified in the given velocity range.\\
{\bf Col. 2}: Velocity range identified from the PV diagrams (See Fig.\,\ref{CO_PV} and Sect.\,\ref{PVmaps}).\\
{\bf Col. 3}: Optical depth of the $^{13}\rm{CO}$  emission calculated using equation (1) from \citet{Shimajiri2014}, assuming an excitation temperature of 10K in local thermodynamic equilibrium (LTE) conditions, and a filling factor of 1. 
{\rev The peak  brightness temperature  of the $^{13}\rm{CO}$ emission over the observed region
 have been derived using the MOMENT task in MIRIAD  for each velocity range.}
%observed 
%in the given velocity range have been derived for 
%is estimated from a Gaussian fitting to the average spectrum %{\rev(shown in Fig.\,\ref{CO_All_spectra})}. 
\\
%{\bf Col. 4}: Column density of the $^{13}\rm{CO}$ emission observed in the given velocity range calculated using equation (2) from \citet{Shimajiri2014}. {\rev Median values of the  $^{13}\rm{CO}$ line width have been derived using the MOMENT task in MIRIAD for each velocity range.}\\
% {\bf Col. 4}: {\rev Median values of the  $^{13}\rm{CO}$ $FWHM$ line width
%derived using the MOMENT task in MIRIAD for each velocity range.}\\
{\bf Col. 4}: {\rev $FWHM$ line width of the  $^{13}\rm{CO}$ spectra
derived using the MOMENT task in MIRIAD for each velocity range.}\\
{\bf Col. 5}: H$_{2}$ molecular gas column density derived from {\rev  $^{13}\rm{CO}$ column densities }
%$N_{^{13}\rm{CO}}$ in the given velocity range
 calculated using equation (2) from \citet{Shimajiri2014},
assuming  a fractional abundance 
of $^{13}\rm{CO}$ with respect to H$_{2}$ of $1.7\times10^{-6}$ \citep{Frerking1982}.\\ 
{\bf Col. 6}: Extend of the emitting structure along the line  of sight estimated from the column density given in {\bf Col. 5} for a critical  density of $2\times10^{3}$\,cm$^{-3}$.\\ 
{\bf Col. 7} to {\bf Col. 10}: Same as {\bf Col. 3} to {\bf Col. 6} for the C$^{18}$O emission, with a  fractional abundance of C$^{18}$O with respect to H$_{2}$  of $1.7\times10^{-7}$   
and a critical density of $5\times10^{3}$\,cm$^{-3}$. \\
%{\rev The column density value of {\bf MF} given in {\bf Col. 9} corresponds to a mean value over the filament. It   is a factor  2  larger when the (peak) column density along the crest is considered.  }
%\\
%{\bf Col. 11}: Gas column density estimated from \herschel\ dust continuum observations averaged over the studied area. \\
$^*$ Median values  for the five structures. % area observed with the Nobeyama 45m telescope. 
\end{tabnote}
\end{table*}

 The velocity fluctuations, as traced  with the C$^{18}$O$(1-0)$   emission along {\bf MF}, are compared to the column density fluctuation as derived from dust continuum using \herschel\ data. 
 We can see velocity gradients towards the cores observed in dust continuum along the filament (Fig.\,\ref{alongfil_3panel}). These may indicate  converging motions towards  the over-densities and prestellar cores identified with \herschel.  The analysis of the velocity structure towards the core is not the scope  of this present paper. 
  Figure\,\ref{alongfil_3panel}-bottom shows the velocity dispersion derived from Gaussian fitting of the optically thin C$^{18}$O$(1-0)$ spectra observed along the crest of {\bf MF}. The total velocity dispersion \citep[cf., e.g.,][]{Arzoumanian2013} is about  $\sim0.27$\,\kms, with some larger values observed at a few positions along the crest of {\bf MF}. We notice an anti-correlation between column density and velocity dispersion, where some of the column density deeps (e.g., around $x_{\rm crest}\sim0.45$) are associated with an increase of the velocity dispersion.

\section{Discussion}\label{Discussion}

In this section, we first discuss the implications of the observed velocity structures towards  {\bf MF}  and {\bf YF}, and their  surroundings in our understanding of filament interaction with the parent cloud gas 
(Sect.\,\ref{filcloud}). Second, we present our observations towards {\bf YF} in the context of  the detection of an early stage of filament formation 
 supported by a theoretical understanding of filament formation in a magnetized ISM (Sect.\,\ref{youngfil}).

\subsection{Filament and sheet-like-cloud interaction}\label{filcloud}

The region studied in this paper is located in the west of the L1495 star forming hub in the Taurus molecular cloud (Fig.\,\ref{Herschel_map}).  A number of starless and prestellar cores are observed along the crest of the main filament {\bf MF} \citep{Marsh2016}. 
This filament has  $M_{\rm line}\sim 16$\,M$_\odot \sim M_{\rm line,crit}$,  a central width of $\sim$0.06\,pc and a power law profile at large radii, {\rev as derived from $Herschel$ observations} \citep[cf. Table\,\ref{tab:FilProp}, Fig.\,\ref{Fil_Herschel} and, e.g.,][]{Arzoumanian2011,Palmeirim2013,Cox2016}.  It has a column density contrast 
with respect to the local background of $\sim 2$ (see Table\,\ref{tab:FilProp} for the adopted definition).
The second filament discussed in this paper, the young filament, {\bf YF}, is thermally subcritical  with  $M_{\rm line}\sim 1$\,M$_\odot << M_{\rm line,crit}$. It has a column density contrast with respect to the local background of $\lesssim 0.3$, a width of $\sim 0.08$\,pc, and its profile is well described by a Gaussian function (Fig.\,\ref{Fil_Herschel}). Table\,\ref{tab:FilProp} summarizes the properties of these two filaments {\rev as derived from $Herschel$ observations}.

These two filaments show coherent, one component velocity structures along their crests, while other velocity components detected towards the filaments are part of more extended structures of the surrounding parent cloud (cf. Sect.\,\ref{Velstruc}). Our results \citep[as opposed to the case of the B211/3 filament, cf.,][]{Hacar2013}  suggest that not all velocity components detected towards {\bf MF} and {\bf YF}  have filamentary, i.e., elongated shape.
{\rev These extended structures may however contribute 1) to the total column density observed in dust continuum towards the filament and 2) to  the power-law wings of the filament radial profile at radii larger than the inner width (cf. Fig.\,\ref{Fil_Herschel}).} Mapping observations at scales larger than the filament width, tracing both dense and low density gas,  are thus necessary to describe accurately the structure of the emitting gas (e.g., filament vs. extended structure). 

 \begin{table}[!h] 
  \tbl{Observed properties of the two studied filaments, as derived from \herschel\ dust continuum observations.\\}  
 { \begin{tabular}{ccccccc}
      \hline
      Filament&$l_{\rm fil}$&\fwhmdec&\ml&$\nhh^0$&$\nhh^{\rm bg}$ & $C^0$ \\ 
      & [pc] & [pc] & [\sunpc]&\multicolumn{2}{c}{[$10^{21}$\nhunit]}& \\  
      \hline
 {\bf MF}& 1 & 0.06$\pm$0.03&16&4.7 &2.4&2.00 \\
 {\bf YF}&0.3 & 0.08$\pm$0.02 &1.2&0.8 & 3&0.27 \\
      \hline
    \end{tabular}}
    \label{tab:FilProp}
\begin{tabnote}
$l_{\rm fil}$: Filament length\\
$\fwhmdec$: Deconvolved $FWHM$-width derived from fitting the mean radial column density profile with a Gaussian function.  \\
\ml: Filament mass per unit length.\\ 
$\nhh^0$: Mean background subtracted column density along the filament crest.\\
$\nhh^{\rm bg}$: Local background column density surrounding the filament.\\
$C^0$: Mean filament column density contrast  with respect to the local background with $C^0=\nhh^0/\nhh^{\rm bg}$. %\citep[see][]{Arzoumanian2018}.
\end{tabnote}
\end{table}

From the analysis of the velocity integrated channel maps (as described in Sect.\,\ref{Velstruc})
we propose a representation of the environment of {\bf MF}, attempting to  constrain the relative positions of the emitting gas structures observed in the  five different velocity ranges, {\bf B}, {\bf F1}, {\bf R1}, {\bf R2}, and {\bf F2}. 
From the observed column densities and densities derived from the detection of molecular emission we estimated  LOS depths of the three extended structures  {\bf SB}, {\bf SR1}, and {\bf SR2} that vary {\rev between $\sim0.2$\,pc and $\sim0.4$\,pc for the $^{13}$CO$(1-0)$ emission and $\sim0.1$\,pc and $\sim0.3$\,pc} for the C$^{18}$O$(1-0)$ emission (cf. Table\,\ref{tab:coldens_tau}). As for the extent of the structures on the plane of the sky we are limited by the coverage of the map of our Nobeyama observations, however a comparison with the  $^{12}$CO data of \citet{Goldsmith2008} indicated that the structures identified in our maps  
cover a larger area {\rev (up to several parsecs)} on the plane of the sky {\rev (Shimajari et al. in prep.)}. The estimated LOS depth suggest that these extended structures are most probably sheet-like (and not spherical). The LOS depths estimated assuming $^{13}$CO$(1-0)$ and C$^{18}$O$(1-0)$ emission associated to densities $>10^3$\,cm$^{-3}$  (cf. Table\,\ref{tab:coldens_tau}) are compatible with the detection of extended structures in $^{12}$CO$(2-1)$, $^{13}$CO$(2-1)$ and C$^{18}$O$(2-1)$  towards our studied field \citep{Tokuda2015}. These latter transitions are shown to  be excited at densities $>10^3$\,cm$^{-3}$ \citep[e.g.,][]{Nishimura2015}.

The PV diagrams show ``velocity bridge" like structures between the velocities of the {\bf SB} and {\bf SR1} structures and that of  {\bf MF} (Fig.\,\ref{CO_PV}). This would suggest a physical connection between these different structures. 
{\rev In the northern section  of {\bf MF}, the correlation between the LOS velocity fluctuations and the column density fluctuations indicate that 
%The observed LOS velocity fluctuations in the North of {\bf MF}, where 
1)} the low column density parts of {\bf MF} have velocities compatible with the {\bf B} velocity range of the extended structure {\bf SB} observed mostly towards the East  of the filament, 2) while the high column density fragments are detected with velocities closer to those observed in the southern, more denser part of the filament. This would suggest that  {\bf SB} %and {\bf SR1}  
may be interacting with {\bf MF}, dragging  along its low column density  parts, while the higher column density fragments are less affected and are observed at
their ``initial" LOS velocities, {\rev i.e, before interaction with {\bf SB}}. 
{\rev In this section of the filament,} we observe a one-sided compression, with  {\bf SB}  sweeping up the low column density parts of  {\bf MF}.  
In the southern section, {\rev  {\bf MF} has 1) velocities in between the {\bf B} and {\bf R1} velocity ranges, 2) smaller velocity fluctuations, and 3) larger column densities.} These observations suggest that   {\bf SB} and {\bf SR1} may be converging simultaneously towards   {\bf MF}, resulting in a compression (column density enhancement) of the filament.
{\rev  The large scale velocity gradient of about 1\kms\,pc$^{-1}$ observed along the filament crest (Fig.\,\ref{vel_alongfil}) may thus result from the interaction of {\bf SB}  and  {\bf SR1}  with {\bf MF}. } 
{\rev Interestingly,   {\bf MF} is observed at similar velocities as that of the neighboring $\sim6$\,pc long  B211/3 filament in the South-East of the field (see Fig.\,\ref{Herschel_map}-left). Using large scale $^{12}$CO and $^{13}$CO observations \citep[from][]{Goldsmith2008}, \citet{Palmeirim2013} identified  velocity gradients on both sides of  the B211/3 filament  with blueshifted and redshifted velocity components with respect to the velocity of the B211/3 filament. This velocity structure is  discussed as tracing a matter flow, confined in a sheet, onto the filament  \citep[][Shimajiri et al. in prep.]{Palmeirim2013}. 
The velocity ranges of the blueshifted and redshifted components observed around the B211/3 filament are similar to that of {\bf SB}  and  {\bf SR1} identified towards  {\bf MF}, supporting our proposed picture of interaction of {\bf SB}  and  {\bf SR1}  with {\bf MF}, % (rather than resulting from, e.g., rotation of the filament), 
 suggesting a coherent picture within the cloud at larger scales.}
%The velocity gradients, that can be seen on the PV diagrams of Fig.\,\ref{CO_PV}, bridging the velocities of {\bf SB}  and  {\bf SR1}  to the velocity of {\bf MF} are unlikely to result from  rotational motions of the entire system (filament and sheets). Indeed, rotation tend to create symmetric structures, while in the studied region the various structures have  different morphologies and spatial extents.  }
%
{\rev The converging motions of {\bf SB}  and  {\bf SR1}  towards  {\bf MF}, % suggested by the  observed  velocity field and spatial distribution of the different structures,
%LOS velocity fluctuations along {\bf MF} and spatial distribution of {\bf SB}  and  {\bf SR1}, 
may also be  in agreement  with the analysis presented in Sect.\,\ref{Ana_12CO}, where 
 %In this latter section 
% we  proposed that 
the decrease  of the  $^{12}$CO intensity 
in the {\bf B} velocity range observed towards {\bf MF}, would result partly from the absorption of the optically thick $^{12}$CO emission  located in the back of {\bf MF} along the LOS. %The {\bf SB}  and  {\bf SR1} structures identified in this field are 

}

This interaction   between the filament and the surrounding more extended (sheet-like) structures may have implications in the evolution and the life time of filaments observed in molecular clouds. Thermally transcritical and supercritical filaments undergoing fragmentation into star forming cores may be affected by the interaction with these surrounding structures. Such interactions, which may vary as a function of relative density and velocity between the filament and the sheet,  may be responsible in sweeping up matter from the filament and the surrounding of the core changing their mass accretion and final total mass. Overall, such interactions may affect/change the star formation activity along the filament.   This may be an example of {\it resetting of star formation activity} along the filaments and might be important in our understanding of the 
 observed star formation efficiency  in molecular clouds.

	    \subsection{Detection of an early stage of filament formation}\label{youngfil}

\begin{figure}
 \begin{center}
  \hspace{-.5cm} 
  \includegraphics[width=9cm]{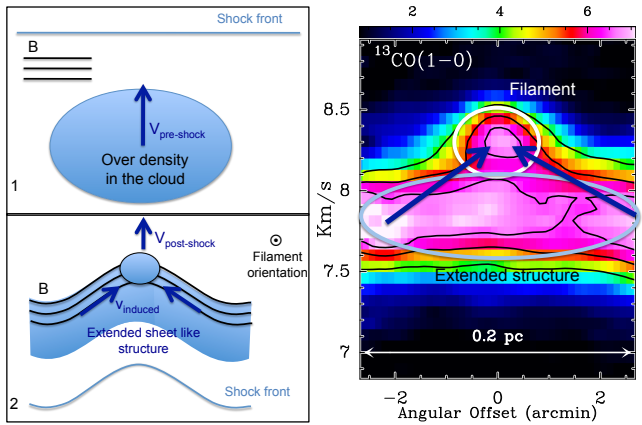}
 \end{center}
          \caption{ {\bf Left:} Schematic view of the filament formation process due to compression of an over density by a shock wave  \citep[adapted from][]{Inoue2017}. the ambient B-field {\rev(indicated with the black lines)} is  parallel to the shock front. 
{\bf Right:} Blow up of the PV diagram (see Fig.\,\ref{PV_ChannelMap_Fil2}) in $^{13}$CO$(1-0)$ towards the filament {\bf YF} for velocities between 6.8\kms and 9\kms.   
	The filament and the parent extended sheet-like structure are indicated with a white  circle and a gray ellipse, respectively. The velocity gradients induced by the bending of the magnetic field are shown as converging arrows towards the filament (cf. Sect.\,\ref{youngfil}).               }
	   \label{PVfil2_SchematicPicture}       
	    \end{figure}

The PV diagrams derived from C$^{18}$O$(1-0)$ and $^{13}$CO$(1-0)$ observations perpendicular to the young filament {\bf YF} (Fig.\,\ref{PV_ChannelMap_Fil2}), 
show  the filament detected at velocities around 8.2\kms, while  a more extended structure is identified at velocities  around 7.8\kms.
The presence of a velocity gradient of about 0.5\kms, between {\bf YF} and the extended  structure, suggests matter flow onto  the young filament, which evolves  increasing in mass and density.  
This observed velocity pattern is in agreement with the formation of filamentary structures by accumulation of matter along magnetic field (B-field) lines bent by a shock compression as proposed by \citet[][see also, Inoue $\&$ Fukui 2013 and Vaidya et al. 2013]{Inoue2017}. 
\nocite{Inoue2013,Vaidya2013}
Appendix\,\ref{AppSimu} presents a PV diagram perpendicular to the main axis of one of the filaments derived from the 3D data cubes of the  simulation by \citet{Inoue2017}.
The schematic view of the filament formation scenario is presented in 
Fig.\,\ref{PVfil2_SchematicPicture} and can be described as the following: \\ 
1. The initial condition corresponds to an over-density,  a clump denser than its surrounding, present in the molecular cloud, formed, e.g., due to interstellar turbulence.  
The clump is threaded with an ordered magnetic field. This structure is about to interact with an interstellar shock front parallel to the magnetic field orientation and propagating  towards it.\\
2. The interaction between the over-density and the propagating shock front compresses the over-dense  structure into a sheet like flattened structure. Due to the finite size of the initial over-dense structure, with respect to the shock front, this latter induces a bent both in the compressed sheet like structure and  in the frozen in  magnetic field structure. The shock front deformation along with the bending of the B-field lines increases  the velocity component along the bent B-field lines, inducing a matter flow along the B-field lines and towards the maximum curvature, where  matter converges as an elongated young filament (see, Sect.\,3.1 of Inoue et al. 2017 for more detailed explanation about this mechanism). Note that in a hydrodynamic oblique shock case velocity gradients are also induced along the oblique shock front, however the presence of an ordered and strong magnetic field is required  to create a long and coherent filament.

Thanks  to the ordered magnetic field structure, and the induced velocity gradients along the B-field lines, the filament is formed perpendicular to the ambiant B-field lines. The filament is thus expected to be straight and  uniform along its crest. The central density and total mass of the filament increase  in time during the evolution  of the sheet-filament system \citep[as suggested by the numerical simulations of][]{Inoue2017}.  

Our observations suggest that matter  converge  along the B-field lines with a rate  of $\rho\,v_{\rm inf}\,2\pi R\approx48$\,\sunpc/Myr, for $\rho=3\times10^3$\,cm$^{-3}$ estimated from  $\nhh=8\times10^{20}\nhunit$ and a size of 0.08\,pc (see Table\,\ref{tab:FilProp}), $v_{\rm inf}=0.5$\,\kms, and $R=0.1$\,pc (see Fig.\,\ref{PVfil2_SchematicPicture}-Right). 
At this estimated rate, the forming filament increases in mass per unit length and may become critical  ($M_{\rm line,crit} \sim16\,\sunpc$) in $\sim0.3\,$Myr. 
 If there is sufficient gas mass in the parent sheet, the initially thermally subcritical young filament may increase in mass per unit length until it reaches a mass per unit length equal to the critical mass per unit length, becomes gravitationally unstable and  undergoes gravitational fragmentation into prestellar cores. On the other hand, when the filament cannot accumulate enough mass  to become gravitationally bound, it may disperse, or be more vulnerable to expected forthcoming collisions/interactions with the surrounding environment.   

Above we compared  the expected velocity pattern from this scenario with the observation of the velocity field  towards and around the filament {\bf YF}. This comparison suggests that  {\bf YF} may be a young filament being formed by the convergence of matter along the B-field lines due to a flow induced by a shock compression. 
The scenario proposed by \citet[][]{Inoue2017} is based on the main role of the relatively strong  magnetic field. 
The only available data tracing the magnetic field towards this region are the \planck\ dust polarization maps at 350\,GHz at a nominal resolution of 5$\arcmin$ \citep[or 0.2\,pc at the distance of 140\,pc of the Taurus cloud, see, e.g.,][]{planck2016-XXXIII}. 
The mean B-field orientation towards {\bf YF} (averaged within $20'\times20'$) is $\sim(112\pm10)^\circ$, making an angle of   $\sim30^\circ$ with respect to the orientation of {\bf YF} on the plane of the sky.   It is however  difficult to conclude on the relative orientation in 3D between the filament and the local B-field with the \planck\  data,  because of 1) the low spatial resolution and the integration of the emission from several structure along the LOS, and 2) the projection effect which may not give the true 3D relative orientation between the B-field and the filament. From  \planck\ dust polarization data we can only derive the plane-of-the-sky (POS) component of the line of sight average B-field. 

Our results suggesting that the B-field structure towards {\bf YF}, a presently subcritical,  low column density filament, being perpendicular to the axis of {\bf YF} may be in contradiction with  the orientation of the B-field lines derived from  \planck\ dust polarization, supporting previous results derived from dust polarization observations as well as near infrared and optical polarization observations \citep[][]{planck2016-XXXII,Soler2016} 
where subcritical filaments are observed parallel to the  B-field lines projected on the POS. 
This discrepancy may  be due to the rapid transition from the thermally subcritical  to the thermally transcritical/supercritical  regimes resulting from the fast ($\sim10\,\sunpc$ in $\sim0.2\,$Myr) accretion of surrounding matter. Such short time scales may be statistically difficult to observe. The observations presented in this paper grasp the {\bf YF} filament at an early stage of evolution, e.g., in the thermally  subcritical stage. We  suggest that {\bf YF} will evolve and become thermally supercritical in the future.  
Molecular line observations towards low column density filament and the analysis of their velocity field combined with the B-field structure would help improving our understanding of the early stages of filament formation. 

\section{Summary and conclusions}\label{Summary}

In this paper,  we presented molecular line mapping observations with the Nobeyama 45m telescope towards a $10'\times20'$ field in the Taurus molecular cloud. 
The analyses and results derived from the OTF maps in $^{12}$CO$(1-0)$, $^{13}$CO$(1-0)$ and C$^{18}$O$(1-0)$, can  be summarized as follows:
 
 \begin{enumerate}
\item 

The C$^{18}$O$(1-0)$ integrated emission traces the elongated structure of the $\sim1$\,pc long  thermally transcritical filament {\bf MF}  with column densities $\nhh\gtrsim5\,\times10^{21}\,\nhunit$ as derived from \herschel\ data. 
%On the other hand 
{\rev The $^{13}$CO and $^{12}$CO integrated intensity maps show more extended emission around  {\bf MF}   with column densities $\nhh\gtrsim1\,\times10^{21}\,\nhunit$. }\\
% between $\sim1$\,and\,4$\times10^{21}\,\nhunit$. \\
  
\item   

Using PV diagrams and velocity channel maps 
derived from  the $^{12}$CO$(1-0)$, $^{13}$CO$(1-0)$ and C$^{18}$O$(1-0)$ emission we identified five structures at the following velocity ranges:  
{\bf B}[4-5.4]\kms, {\bf F1}[5.4-6.2]\kms, {\bf R1}[6.2-7.2]\kms, {\bf R2}[7.2-8.1]\kms, {\bf F2}[8.1-8.7]\kms.
The structures emitting at the velocities of {\bf B},  {\bf R1}, and {\bf R2} are identified as extended sheet-like structures surrounding the filament  {\bf MF}. 
   The {\bf F2} velocity range corresponds to a thermally  subcritical filament ({\bf YF}) detected in the $^{13}$CO channel map.\\  

\item 

We compared the optically thick   $^{12}$CO$(1-0)$ spectra  ``on" and ``off"  {\bf MF}. In the {\bf B} velocity range, we identify  a decrease of $^{12}$CO  intensity ``on" {\bf MF}, that may results from absorption, suggesting that the  structure emitting at these velocities may be located, at least partly, in the back of {\bf MF} with respect to the LOS. Similar analysis suggested that the structures emitting at velocity ranges of  {\bf R1} and {\bf R2} may be located in the front of  {\bf MF}  with respect to the LOS. We use these analysis to describe the relative positions of the sheet-like structures surrounding the filament {\bf MF}. \\

\item 

We detect a velocity gradient along the crest of  {\bf MF}   with velocity oscillations of $\sim0.2\,$\kms\ on average,  increasing towards the North part of the filament. We compare the velocity and column density  structures along the filament crest.  
{\rev In the South part {\bf MF} has 1) larger column densities, 2)  velocities in between that  of {\bf SB} and {\bf SR1}, and 3) small velocity fluctuations. These suggest that   {\bf MF} may be compressed on both  sides by {\bf SB} and {\bf SR1}.
In the northern part,  
 the velocities observed towards the low column density sections   of {\bf MF}} are compatible with that of the {\bf SB} structure emitting at the {\bf B} velocity range. We suggest that, in the northern section,  {\bf SB}  is interacting with  {\bf MF} dragging its low column density parts. The higher column density fragments resist the sweeping and are observed at the LOS velocity of {\bf MF} before interaction with {\bf SB}.\\  

\item 

The PV diagrams of $^{13}$CO and C$^{18}$O towards {\bf YF} show an elongated structure at velocities that differ by $\sim0.5$\,\kms\ with respect to the velocity of {\bf YF}. We compare  the observed velocity structure with that expected from the filament formation model presented in \citet[][see also Appendix\,\ref{AppSimu}]{Inoue2017}. We suggest that our observations are compatible with the formation of a filament from the accumulation of matter along magnetic field lines induced by a  shock compression due to a propagating wave. The observations suggest that a {\bf YF}-like filament may form at a mass accretion rate of $\sim50$\,\sunpc/Myr and become thermally critical in $\sim0.2$\,Myr.\\

\item  

The propagation of interstellar shock waves, creating sheet-like molecular gas structures may play an important role in the formation of filamentary structure in molecular clouds. These filaments increase in mass per unit length accreting matter from the surrounding sheet until reaching the critical mass per unit length and becoming gravitationally unstable, fragmenting into star forming cores.    
The same propagating shock waves may interact with already formed filaments resulting in the compression of additional matter onto the filaments or the removal/disruption of low column density parts.  
We suggest that such interactions may play an important role in the life time of filaments and 
 their star formation activity.  
 
\end{enumerate}

%-------------------------------------------------------------------

%%%%%%%%%%%%%%%%%%%%%%%%%%%%%%%%%%%%%%%

\begin{ack} 
We thank Kazuki Tokuda for discussions  about the CO$(2-1)$ observations of the Taurus molecular cloud.
DA acknowledges an International Research Fellowship from the Japan Society for the Promotion of Science (JSPS).  
The 45-m radio telescope is operated by Nobeyama Radio Observatory, a branch of the National Astronomical Observatory of Japan. 
The numerical computations were carried out on XC30 system at the Center for Computational Astrophysics (CfCA) of National Astronomical Observatory of Japan.
This work is supported by Grant-in-aids from the Ministry of Education, Culture, Sports, Science, and Technology (MEXT) of Japan (15K05039 and 16H02160).

\end{ack}

\appendix 
\section{Column density structure derived  from \herschel\  observations}\label{HerschelObs}
\begin{figure}
 \begin{center}
    \hspace{1cm}  
  \includegraphics[width=7cm]{./Herschel_coldens_skel_2fil.pdf} 
   \hspace{0.5cm} 
  \includegraphics[width=7cm]{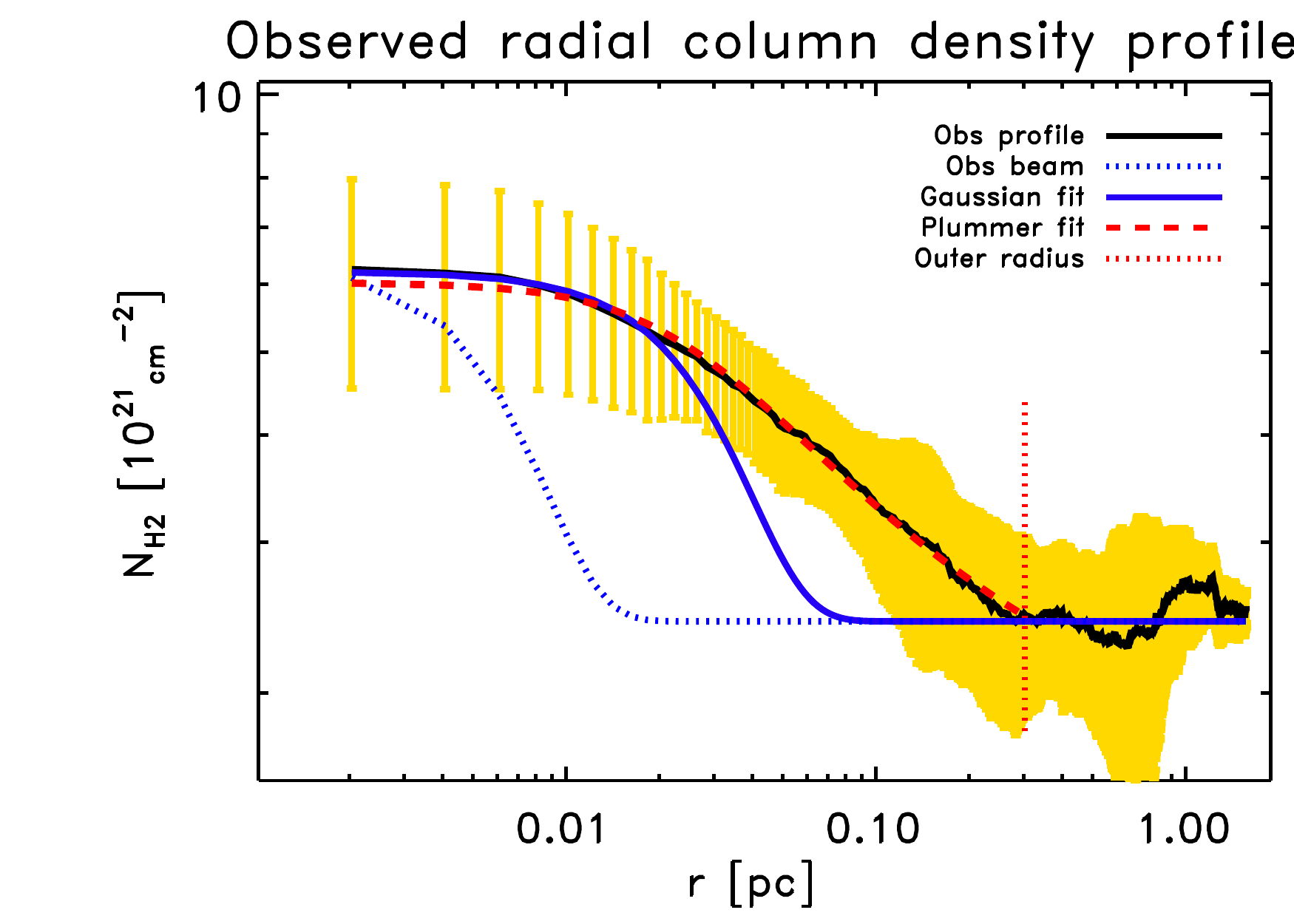} 
    \includegraphics[width=8cm]{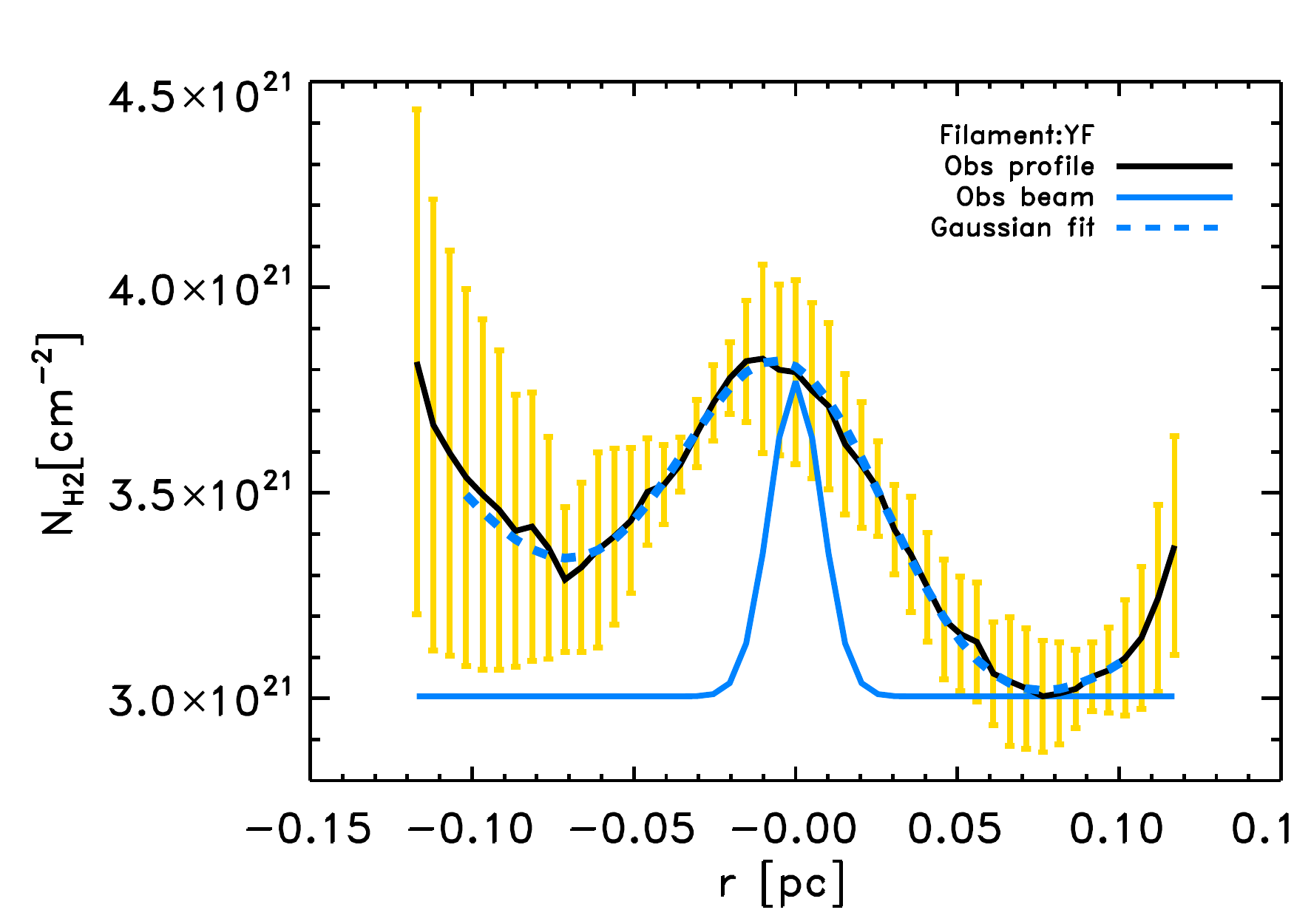} 
 \end{center}
\caption{    {\bf Top:} Blow up  of  the \herschel\   column density (\nhh) map  (at the resolution of 28\arcsec)   
	           towards the region studied in this paper (same as Fig.\,\ref{Herschel_map}). The blue and red skeletons show the crest of the filaments  {\bf MF} and {\bf YF} traced with \disperse, respectively. 
	           The {\bf MF} filament is traced on the column density map, while the {\bf YF} filament has been traced on the $^{13}$CO$(1-0)$ {\bf F2} channel map (Fig.\,\ref{PV_ChannelMap_Fil2}-right). The white dashed contour shows the footprint of the CO maps observed with the Nobeyama 45m telescope.
	            {\bf Middle:} Column density  radial profile (in log-log) perpendicular to the main axis {\bf MF} and  averaged along its crest. The blue and red curves show the best Gaussian and Plummer-like function fits, respectively. The vertical dotted red line marks the outer radius of the filament. The blue dotted curve shows the observational beam (HPBW=28\arcsec). 
	             {\bf Bottom:}    
	             Column density radial profile (in lin-lin) perpendicular to the main axis of {\bf YF}. The best Gaussian fit and the observational beam are shown in dashed and solid blue curves, respectively. Table\,\ref{tab:FilProp}  summarizes the properties of {\bf MF} and {\bf YF} derived from  \herschel\ observations. 
	               }
	   \label{Fil_Herschel}         
	    \end{figure}

We present here the properties of the two filaments discussed in this paper as derived from \herschel\ dust continuum observations. The column density map shown in Fig.\,\ref{Fil_Herschel} is derived as explained in 
 \citet[][]{Palmeirim2013} \citep[see also, http://gouldbelt-herschel.cea.fr/archives, and, e.g.,][]{Marsh2016}. We convolve the column density map to  the $28\arcsec$ resolution, the same as that of the molecular line maps studied in this paper. 
 
 We trace the crest of the filaments using the \disperse\ algorithm  \citet[][]{sousbie2011}.
The {\bf MF} filament is traced on the column density map, while the {\bf YF} filament is traced on the $^{13}$CO$(1-0)$ {\bf F2} channel map (Fig.\,\ref{PV_ChannelMap_Fil2}-right).
We derive radial column density profiles perpendicular to the filament crests and measure  the filament properties as explained in 	 \citet{Arzoumanian2011,Arzoumanian2018}.

  Figure\,\ref{Fil_Herschel} shows the column density map derived from  \herschel\  observations and the radial column density profiles perpendicular to the  {\bf MF}  and  {\bf YF}  filaments and averaged along their crests. Table\,\ref{tab:FilProp} summarizes the main properties of the filaments.

\section{Analysis of the observed spectra over the field}\label{AppSpectra}

\begin{figure}
 \begin{center}
   \hspace{-0.9cm}  
 \includegraphics[width=9cm]{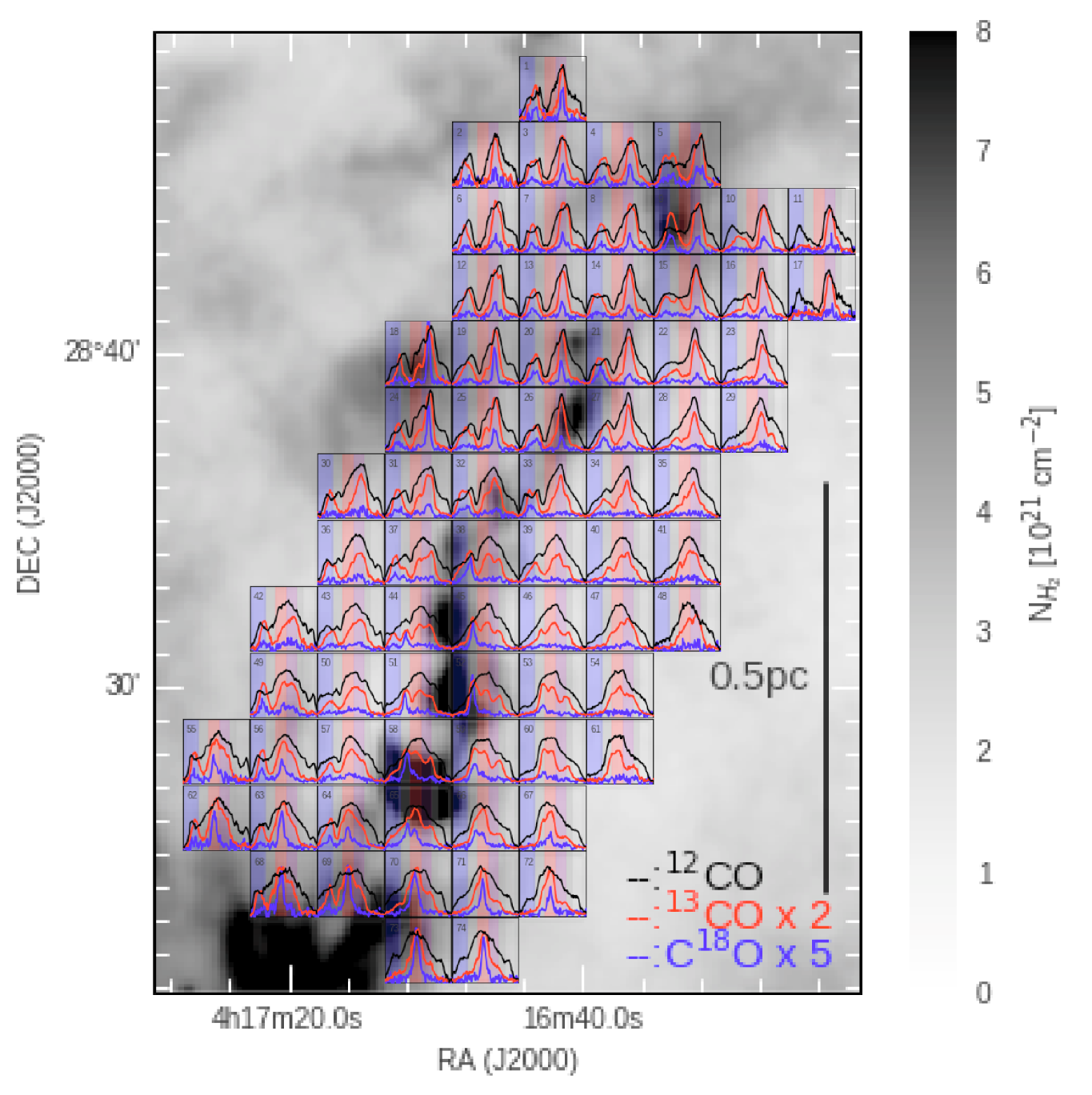} 
 \end{center}
  \vspace{-0.3cm} 
\caption{   \herschel\   column density  map  at the resolution of 28\arcsec (same as Fig.\,\ref{Herschel_map}), overlaid with spectra averaged in $2'\times2'$ squares over the observed region. The velocity range is [4-10]\kms, and the intensity range  is from -1 to 18\,K. 
Black, red, and blue spectra correspond to $^{12}$CO$(1-0)$, $^{13}$CO$(1-0)$, and C$^{18}$O$(1-0)$, emission respectively. The  $^{13}$CO line is multiplied by a factor of 2 and the C$^{18}$O line  by a factor of  5.  The shaded area on the plots for each spectrum show the same velocity ranges as the 5 channels discussed in the paper (see, e.g., Fig. \,\ref{COspectra_map}). 	               }
	   \label{SpectraOnMap}         
	    \end{figure}

In this appendix we extend the analysis discussed in Sect.\,\ref{Ana_12CO} presenting averaged $^{12}$CO$(1-0)$, $^{13}$CO$(1-0)$, and C$^{18}$O$(1-0)$ spectra overlaid on the \herschel\   column density map (see  Fig.\,\ref{SpectraOnMap}). The spectra have been averaged over  $2'\times2'$ squares and centered at different location over the observed field on the filament {\bf MF} as well as on its East and West sides. Figure\,\ref{SpectraOnMap}, complements the spectra of  Fig.\,\ref{COspectra_map} showing the variation of the spectra in the {\bf B} velocity range suggesting possible absorption of the optically thick emission, 
while such features are not observed towards the {\bf R1} and {\bf R2} velocity ranges. We have used this analysis to suggest a 3D structure of the observed portion of the cloud (see Sect.\,\ref{Discussion}). 

%\appendix
\section{Position-velocity diagram derived from  the numerical simulation of filament formation by \citet{Inoue2017}}\label{AppSimu}

In this section we present the column density map and the PV diagram towards a filament extracted from the numerical simulation of \citet{Inoue2017}, which studied the formation and evolution of filamentary structures 
 induced by a shock propagation in a turbulent molecular cloud. The numerical simulation 
  is an isothermal MHD simulation using an  adaptive mesh refinement technique.
The simulation has been particularly tuned to study the formation of massive filaments forming high mass stars, thus the densities and the velocity of the shock propagation are extreme compared to the low mass star forming region such as the Taurus molecular cloud, however here we  confirm that  the filament observed in the simulation is formed by the mechanism explained in Sect.\,\ref{youngfil}, and has a velocity structure similar to that of the filament {\bf YF} (see right hand side panel of Fig.\,\ref{PVfil2_SchematicPicture}).

\begin{figure}[h]
\includegraphics[scale=0.45]{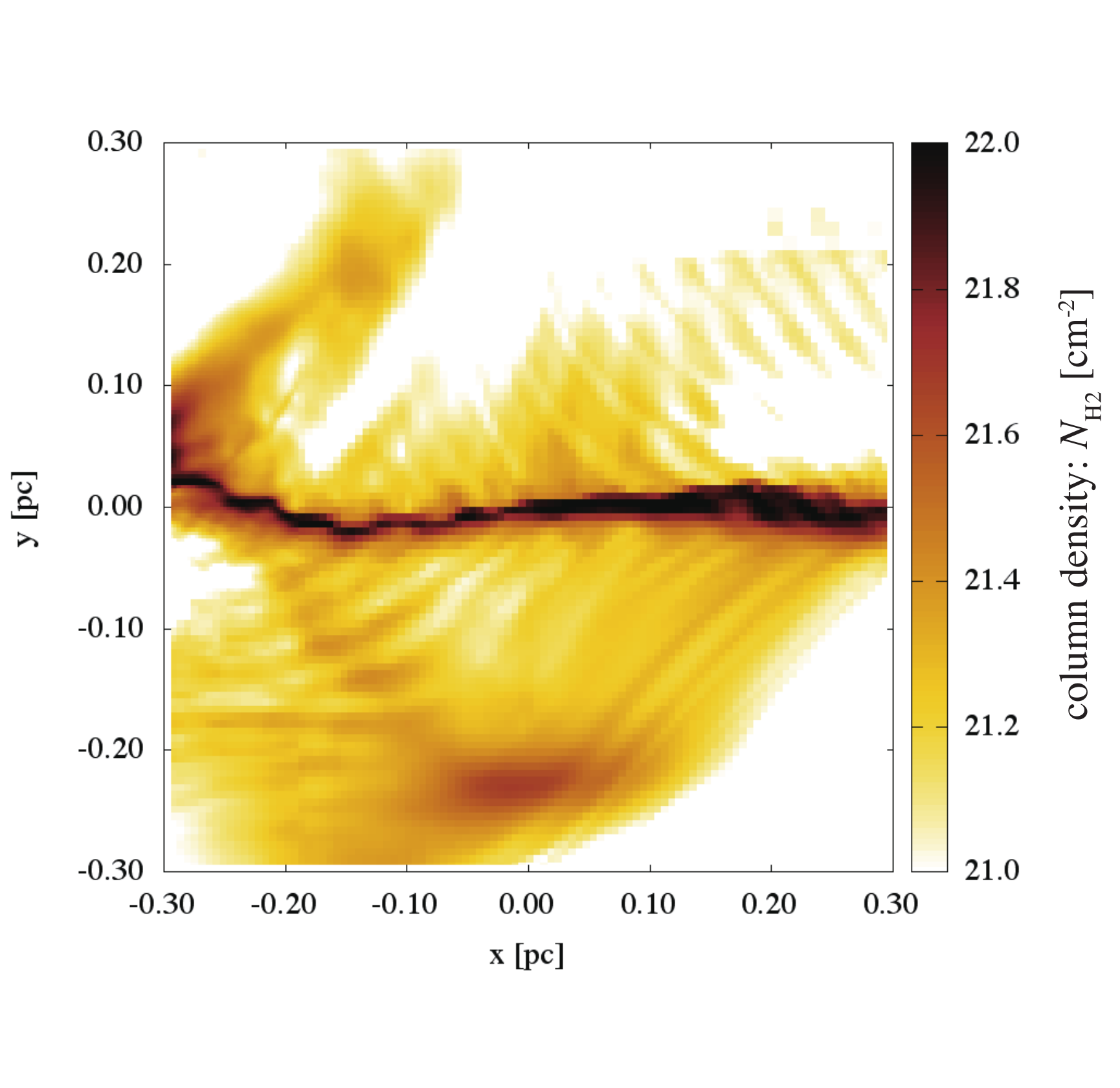}
\caption{\label{fa1}
Background subtracted column density structure of a filament from the simulation data by \citet{Inoue2017}.
}\end{figure}

Figure \ref{fa1} shows the background subtracted column density structure of a filament from the simulation of \citet[][]{Inoue2017}  identified to be formed by the mechanism explained in Sect.\,\ref{youngfil}.
We have selected a snapshot data (at $t=0.29$ Myr after the shock compression) before the filament evolves into supercritical and shows gravitational collapse (that happens at $t=0.44$ Myr).
The mass per unit length  of the filament at this epoch is $\sim 15\,$M$_{\odot}$/pc$^2$, which is larger than the filament {\bf YF}, but we confirmed that this filament is growing in mass 
as illustrated in the bottom left panel of Fig.\,\ref{PVfil2_SchematicPicture}.

\begin{figure}[h]
\includegraphics[scale=0.45]{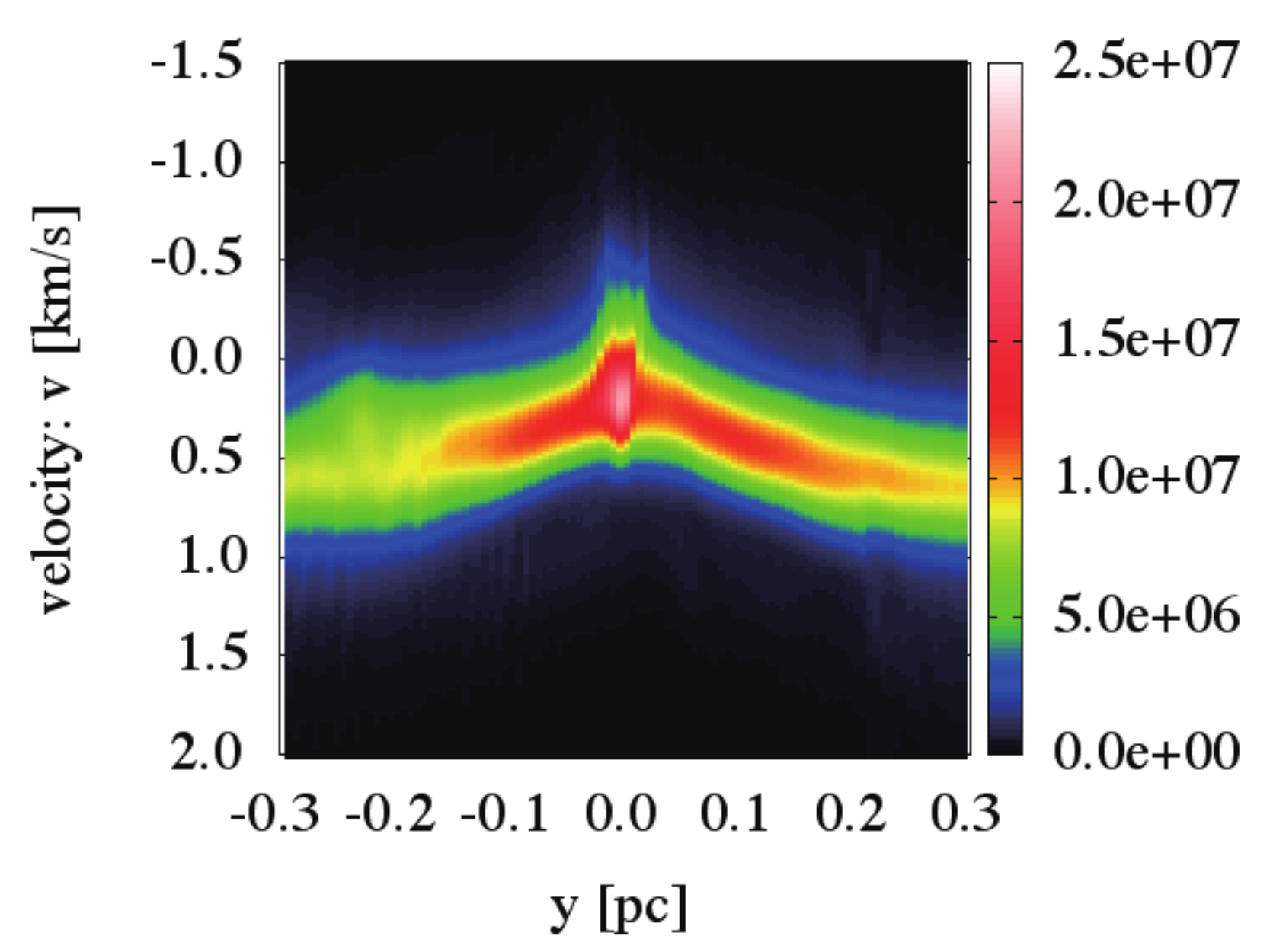}
\caption{\label{fa2}
Position-velocity diagram perpendicular to the filament shown in Fig.\,\ref{fa1}  from the simulation by \citet{Inoue2017}. This diagram is derived by averaging the simulated cubes along the filament length as described   
by  Eq.\,(\ref{eq:pv}). The unit of the color scale of the map is given by Eq.\,(\ref{eq:pv}).
}\end{figure}

To obtain a synthetic position-velocity diagram across the filament and averaged along the $y$-axis (Fig.\,\ref{fa1}), we apply the following mapping formula to the 3D simulation data cube:
\begin{equation}\label{eq:pv}
F(y,v)=\int \int\,\rho(x,y,z)\,\exp [ -\{v-v_{z}(x,y,z)\}^2/c_{\rm s}^2]\,dx\,dz,
\end{equation}
where we assume $c_{\rm s}=0.2$ km/s.
The resulting map is shown in Fig.\,\ref{fa2}.
We see that the structure resembles that of the filament {\bf YF} shown in the right panel of Fig.\,\ref{PVfil2_SchematicPicture}.

\newpage
\bibliographystyle{aa}
\bibliography{aa}

\end{document}